# Regulation of Macrocycle Shuttling Rates in [2]Rotaxanes by Amino-Acid Speed Bumps in Organic–Aqueous Solvent Mixtures

Maxime Douarre,[a] Vicente Martí-Centelles,*[a] Cybille Rossy,[a] Isabelle Pianet,[b] and Nathan D. McClenaghan*[a]

[a]  M. Douarre, Dr. V. Martí-Centelles, Dr. C. Rossy, Dr. N. D. McClenaghan
    Institut des Sciences Moléculaires,
    CNRS UMR 5255, University of Bordeaux,
    33405 Talence, France.
    E-mail: nathan.mcclenaghan@u-bordeaux.fr
[b]  Dr. I. Pianet
    IRAMAT, CNRS UMR 5060, Maison de l'Archéologie, Université Bordeaux Montaigne,
    33607 Pessac, France.

    Supporting information for this article is given via a link at the end of the document.

**Abstract:** A homologous series of two-station [2]rotaxanes incorporating amino-acid units in the molecular thread has been developed. The degenerate [2]rotaxanes exhibit amino-acid specific shuttling rates between two fumaric stations related to the steric factor associated to the amino-acid side chain, as demonstrated by variable-temperature $^1$H NMR spectroscopy and exchange spectroscopy (EXSY). This allows tuning of the macrocycle shuttling rate over 4 orders of magnitude, which further has a relatively small solvent dependency.

## Introduction

Rotaxanes are the most commonly used systems for constructing artificial molecular machines.[1] Indeed a variety of effective methods for the synthesis of rotaxanes has been established through numerous synthetic strategies.[2–5] Among a wide diversity of reported interlocked structures, molecular shuttles comprising a mobile macrocycle threaded on a two-station axle, is one of the most emblematic types of architecture, which serves as an excellent platform for bottom-up studies of dynamic properties.[6–8] With sub-molecular movement being integral to potential function in molecular machines, regulation of dynamics, both spontaneous and stimulus-induced, is essential and intimately linked with molecular design.[9–12] Developing methods to control the shuttling rate could unlock pioneering applications in the field of catalysis, as exemplified by rotating catalysts where the rotation speed had a direct influence in their performance.[13]

Despite the central importance of the ring shuttling rate in rotaxanes, only a relatively small number of studies have focused on detailed kinetic analyses providing key structure-property information.[14–16] Relevant works include the discovery of the key role of water lubricating hydrogen-bonded molecular machines by Leigh,[17] Stoddart,[18] and Harada.[19] The development of systems with steric barriers for unidirectional movement by Stoddart,[6] and the null impact of the rotaxane axle length on the ring shuttling reported by Hirose.[14] These contributions are central to understanding the effect of the different macrocycle-thread interactions on the shuttling rate and incorporate such structural parameters in the design of rotaxanes with specific shuttling dynamics.

A family of interest are Leigh's rotaxanes, assembled in one synthetic step and facilitated by hydrogen bond pre-organisation evoking a tetraamide macrocycle and fumaramide or succinamide station on the thread.[20–23] In these systems, polar solvents have a significant effect on the shuttling rate, especially water,[17,24] producing an increase of the shuttling rate as a consequence of disrupting hydrogen bonding between the macrocycle and the station, on providing competitive H-bonding interactions with the solvent. Despite the easy synthetic accessibility, the control of shuttling rate of the macrocycle has not been well-established and is highly dependent on the polarity solvent. Achieving this control in aqueous solvent mixtures, and ideally in water, is needed for developing operational rotaxanes for potential biological applications. Yet, few examples of rotaxane shuttles in water have been reported with limited ring shuttling kinetic information.[25–28] While water is shown to accelerate ring movement in hydrogen-bonding rotaxanes, full control implies the ability to slow ring movement and even render it solvent independent to a large degree. For this purpose, introducing a custom steric barrier on the thread that the macrocyclic can only pass over slowly, can produce the desired shuttling rate reduction.

The selection of the steric barrier size is critical, as just one methyl difference may completely block the ring shuttling, as highlighted by macrocycle dethreading experiments in rotaxanes with stoppers of different sizes.[29] In this work we developed designer rotaxane systems building on a benchmark system, namely Leigh's fumaramide rotaxane (Figure 1 top), by incorporating steric barriers to control the macrocycle shuttling rate in both non-polar organic solvents and organic aqueous mixtures.[30,31] For this propose, we synthesized degenerate molecular shuttles containing specific amino-acid units in the thread which are able to act as speed bumps to control the macrocycle shuttling speed (Figure 1, bottom) and sought to generalize this approach for a range of amino-acid units. In these systems, which have two identical stations, the macrocycle would have an equal probability of residing on each station and stochastic macrocycle shuttling would be operational. These degenerate [2]rotaxanes systems provide an appropriate platform to measure the shuttling process by dynamic NMR spectroscopy using variable temperature $^1$H NMR and EXSY NMR.

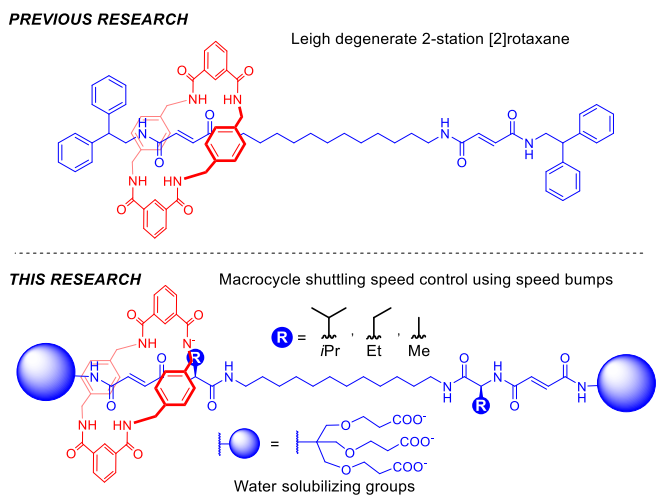

**Figure 1.** Representation of Leigh's degenerate 2-station [2]rotaxane (from Ref [21]) and the analogues developed in this project incorporating amino-acid speed bumps and water solubilizing groups on the stopper.

## Results and Discussion

The prototype 2-station molecular machine design developed by Leigh and coworkers (Figure 1 top),[20] which comprises a tetraamide macrocycle and two-fumaramide stations, has a typical energy barrier of 16.2 kcal/mol to ring displacement corresponding to a shuttling speed of 8 times per second in $CDCl_3$ at 25 ºC.[21] This barrier is greatly affected by polar solvents such as water that weaken the hydrogen-bond strength between the macrocycle and the station. This results in a reduction of the energy barrier and consequently an enhancement of the shuttling rate.[17,24] In the current design (Figure 1 bottom), we integrated a multi-ester stopper that can be hydrolyzed to obtain six negatively charged carboxylate groups offering enhanced solubility in polar solvents, including water. Additionally, amino-acid units were integrated into the molecular thread to act as speed bumps in our prototype rotaxane model (Figure 1, bottom). In particular, amino-acid speed bumps with R = Me, Et and *i*Pr (using the amino acids Ala, Abu and Val, respectively) have an ideal size to act as steric speed bumps for the macrocycle employed. The wide variety of commercially-available natural and non-natural amino acids affords a selection of side chains with the appropriate size and simplifies the synthesis of the corresponding axles. Among different possible stations, the fumaramide station was selected as it has the larger hydrogen bonding affinity towards the macrocycle it favours the positioning of the macrocycle over the station in polar solvents. Whereas the strongly hydrogen bond disrupting solvent DMSO displaces the macrocycle from the low affinity dipeptide stations,[32] the fumaramide station is still occupied by the macrocycle for at least 85% of the time in DMSO.[33] Stopper **6** (Figure 2) was designed to impart water solubility of the rotaxanes in water upon hydrolysis of the tert-butyl groups. Furthermore, the tert-butyl groups enhanced the solubility of thread **2** and rotaxane **1** in organic solvents.

The synthesis of these structures was carried out as described in Figure 2. Amide bond formation on reacting 1,12-diaminododecane with the activated Boc-Aaa-OH followed by deprotection yielded compound **3** in 80–90% yield over two steps. Stopper **6** was prepared according to a literature procedure by condensation of tris(hydroxymethyl)aminomethane (TRIS) with tert-butyl acrylate.[34] Amide bond formation with monomethyl fumarate followed by deprotection in basic conditions yielded **4** in 61% yield over two steps. Peptidic coupling of **3** with **4** yielded thread **2** in 40–50% yields. The final rotaxane formation reaction was performed by simultaneously separate addition of chloroform solutions of p-xylylenediamine and of isophthaloyl dichloride to that of thread **2** and triethylamine in chloroform during 2 h via a syringe pump. The target rotaxanes **1** were successfully isolated in 10–16% yield after column chromatography purification. Hydrolysis of rotaxane **1a** in neat formic acid yielded rotaxane **1a-WS** quantitatively. The rotaxanes were characterized by $^1$H, $^{13}$C, 2D NMR and high resolution mass spectrometry.

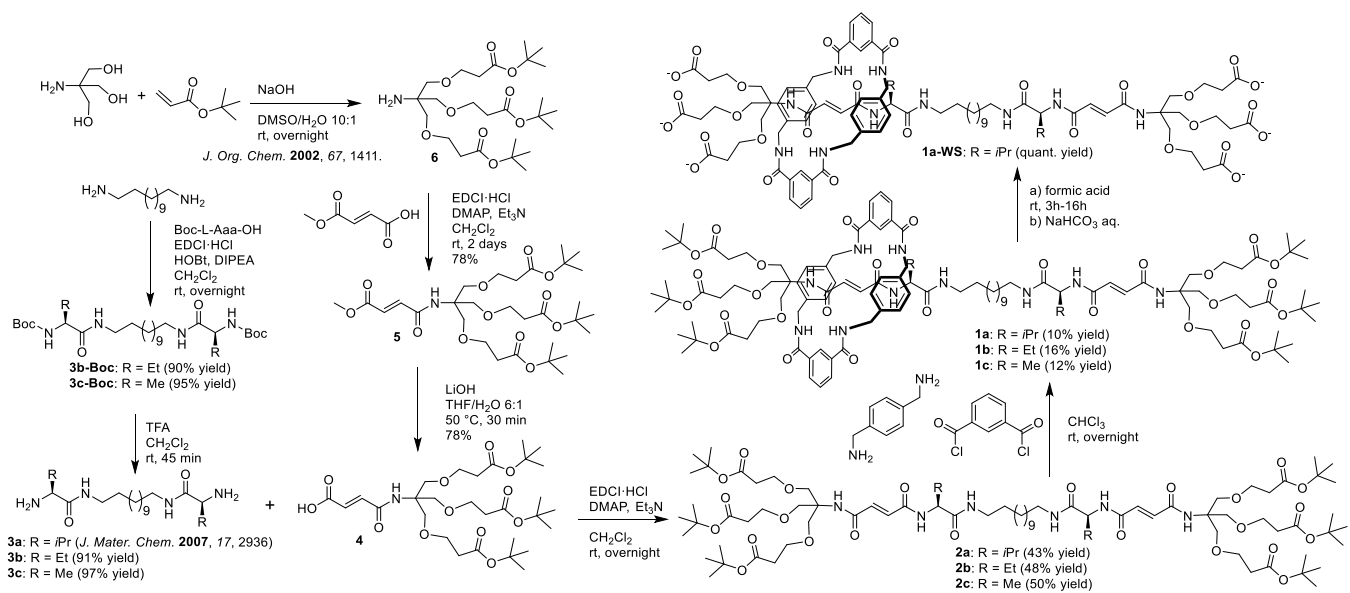

**Figure 2.** Synthesis of the [2]rotaxanes. Stopper **6** and valine amide **3a** were prepared using literature procedures.[34–36]

Initial analysis of the $^1$H NMR of rotaxane **1a** in CDCl$_3$ at room temperature showed two sets of signals for the fumaramide protons, which could be assigned to the free and occupied stations. Whereas the free station fumaramide proton resonances (S2) had the same chemical shift observed in the thread, the presence of the macrocycle produced a Δδ > 1 ppm upfield shift in the fumaramide signals S2 attributed to the deshielding produced by the macrocycle phenyl rings (Figure 3). The dynamic nature of ring movement within the rotaxane was determined by EXSY $^1$H NMR. EXSY allows direct measurement of dynamic process in equilibrated systems as cross peaks between the two signals that undergo exchange during the mixing time of the experiment.[37]

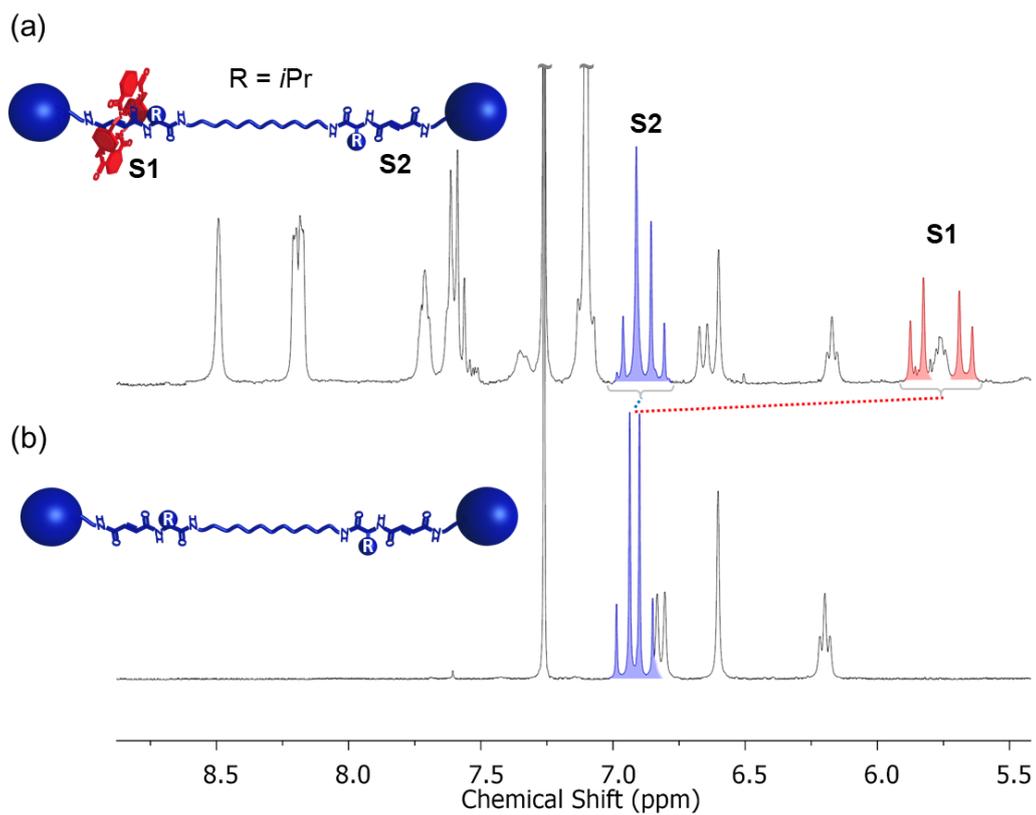

**Figure 3.** $^1$H NMR in CDCl$_3$ at rt, 300 MHz: (a) Rotaxane **1a**, (b) thread **2a**. Fumaramide protons labelled as S1 for occupied station and S2 for free station.

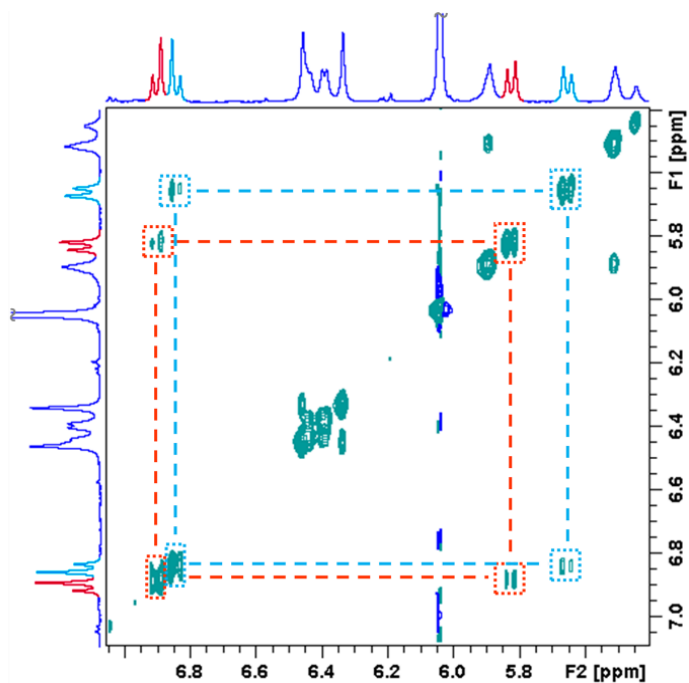

**Figure 4.** EXSY $^1$H NMR of rotaxane **1a** in TCE-$d_2$ at 85 °C, 600 MHz, 250 ms mixing time.

Observation of cross peaks between the two fumaramide protons confirmed the chemical exchange of signals S1 and S2, which is consistent with spontaneous macrocycle shuttling between the two similar stations (Figure 4).

A similar spectrum was obtained for rotaxane **1b**, comprising an ethyl appendage rather than an isopropyl group, two sets of signals at room temperature for the fumaramide protons being observed. In contrast, further downsizing to a methyl appendage in rotaxane **1c** resulted in a single set of broad signals for the fumaramide protons, indicating a relatively fast exchange between the signals of the two stations on the NMR timescale (Figure 5). This observation qualitatively confirms that the macrocycle undergoes a rapid shutting in this case.

To evaluate the effect of the amino-acid speed bump on the macrocycle shuttling, variable temperature (VT) $^1$H NMR experiments in TCE-$d_2$ were performed (600 MHz spectrometer). Whereas no signal coalescence was observed for rotaxane **1a** in the 25–105 °C temperature range, we could observe coalescence for rotaxanes **1b** and **1c**. This showed spontaneous shuttling on the ms–μs timescale at higher temperature. The *t*Bu signals ($\Delta\delta \approx$ 0.045 ppm) of the stopper has a coalescence temperature > 105 °C for **1a**, 75 °C for **1b** and –10 °C for **1c**, showing a direct effect of the steric bulk on the shuttling rate, the larger the substituent (*i*Pr > Et > Me) the lower the rate. A similar trend was observed for the fumaramide signals ($\Delta\delta \approx$ 1 ppm). In this case we only observed a coalescence of the fumaramide signals for rotaxane **1c** (Figure 5). Although no coalescence was reached for the fumaramide signals in rotaxanes **1a** and **1b**, signal broadening indicated the slow dynamics of the macrocycle shuttling in these systems. Overall, the differences observed in coalescence temperatures reflect a reduction of the ring shuttling rate by the increase in amino-acid unit steric size from **1c** to **1a** (Table 1).

**Table 1.** Coalescence temperatures in TCE-$d_2$, 600 MHz. Signal width at half-height in brackets at the temperature displayed in the table for protons not reaching coalescence. Fumaramide signals Δδ ≈ 1 ppm (600 Hz). *t*Bu signals Δδ ≈ 0.045 ppm (27 Hz).

| Rotaxane | Fumaramide | Stopper *t*Bu |
|---|---|---|
| **1a** | > 105 °C (7 Hz) | > 105 °C (8 Hz) |
| **1b** | > 95 °C (20 Hz) | 75 °C |
| **1c** | 25 °C | –10 °C |

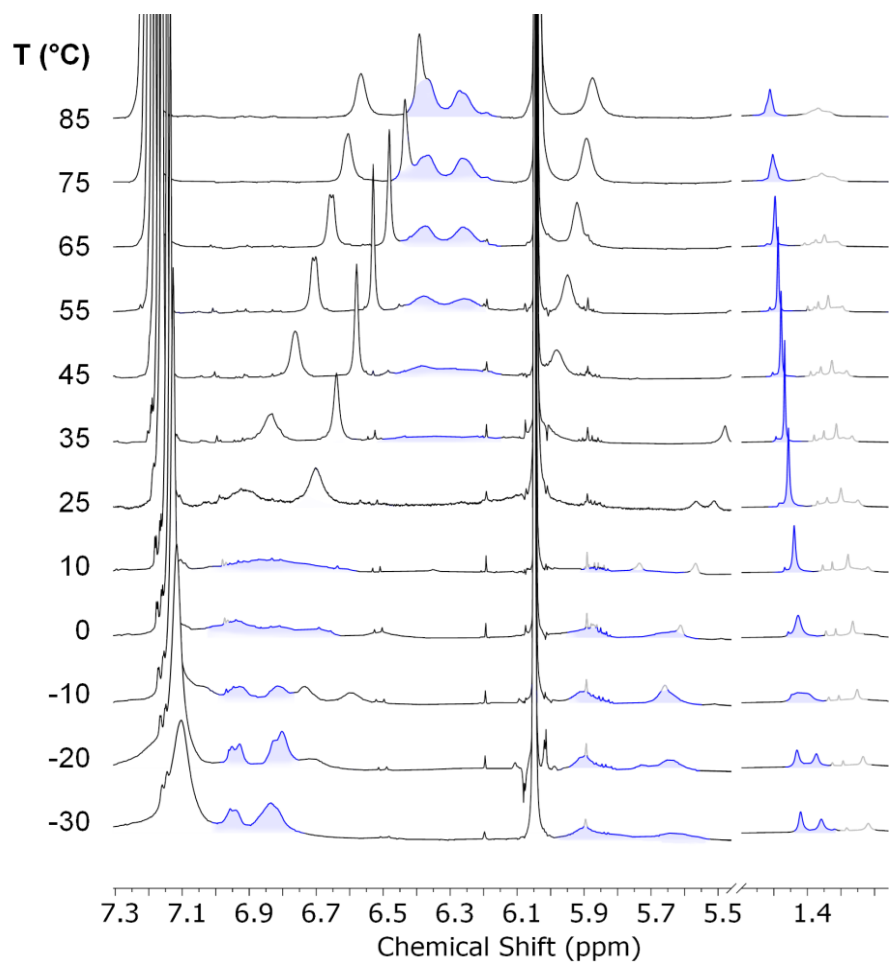

**Figure 5.** VT ¹H-NMR of alanine-containing rotaxane **1c** with the signals displaying coalescence marked in blue (600 MHz in TCE-$d_2$). Solvent impurities are marked in grey.

From the VT ¹H NMR experiments it is possible to obtain further information on the dynamics of the process. Increasing the temperature produces an increment of the exchange rate, producing a broadening of the signals until coalescence is achieved when the exchange rate approaches the frequency difference between the two signals. Further heating produces a fast exchange and a spectrum with individual signals at the average frequency of the signals in this symmetrical molecule. This phenomenon can be used to extract quantitative information of the dynamic process by performing a dynamic NMR lineshape fitting.[14,38,39] The results show that the amino-acid speed bump has a direct influence in the shuttling rate, spanning the measured $k_{ex}$ over 4 orders of magnitude (Figure 6 and Table 2). For the three rotaxanes **1a**, **1b** and **1c**, EXSY ¹H NMR experiments showed cross peaks connecting the different fumaramide and the *t*Bu protons allowing determination of the associated $k_{ex}$ (see Figure 6 and supporting information for further details).

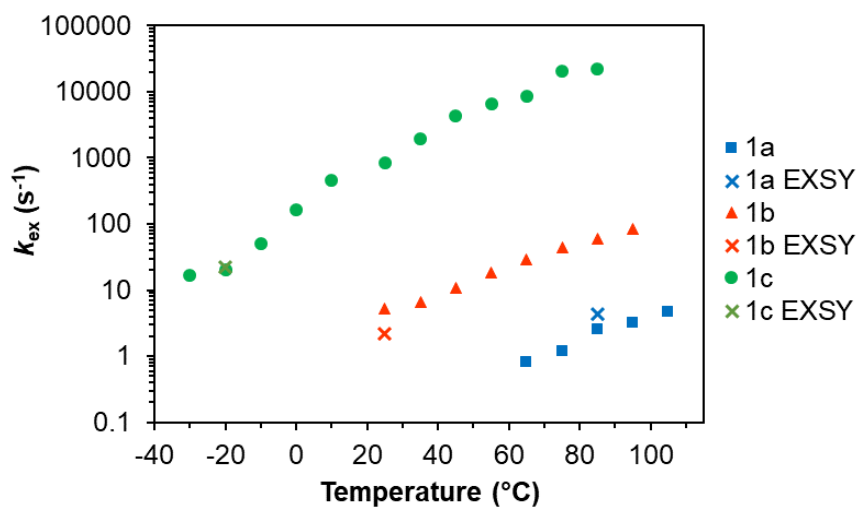

**Figure 6.** Exchange rate constants determined for rotaxanes **1a**, **1b** and **1c** in TCE-$d_2$ at different temperatures by lineshape analysis (squares, triangles and circles) and EXSY (crosses) $^1$H NMR, 600 MHz.

In order to rationalize the significant changes in the shuttling rates introduced by the amino-acid speed bumps, we evaluated the precise steric hindrance introduced in each case. To this end, we used the steric parameter $S^0$ of the Me, Et and *i*Pr groups to determine if there is a direct relation between sterics vs energy barrier.[30,40] The steric parameter $S^0$, derived empirically from the quaternization rate constant of methylation reactions of ortho-substituted pyridines, has been proposed as a quantitative scale in the evaluation the steric bulk of substituents in the shuttling behavior of [2]rotaxanes.[30] A plot of $\Delta G^\ddagger$ vs $S^0$ shows a linear relation showing the direct relation and preponderant nature of the steric parameter on the process. This data suggests that the speed bump does not significantly affect the complementary hydrogen bonding interactions between the tetraamide macrocycle and the fumaramide stations. Therefore, the energy levels of S1 and S2 would remain unchanged and the speed bumps only affects the energy of the transition state that connects stations S1 and S2 as schematized in Figure 7b.

We further considered the CPK substituent volume as an alternative parameter that can be used to evaluate the steric bulk of amino-acid side chain.[41] The CPK volumes offer a good linearity with the observed experimental shuttling energy barrier (see Figure 7d and Supporting Information). We envisage that CPK volumes could be a useful tool to select appropriate speed bumps in rotaxane systems. As cited above, the parameter $S^0$ provides a good description of the overall steric bulk in systems for small groups such as Me, Et and *i*Pr groups. However, $S^0$ does not describe accurately the overall steric bulk of larger groups, while the use of the relative CPK volume of the amino-acid side chain with regard to the macrocycle cavity size provides a better description of the speed bump steric size (see Supporting information). Analysis of the macrocycle cavity showed a diameter of 4.88 Å, corresponding to a spherical void volume of 61 Å$^3$. The overall size of the speed bump was obtained as the CPK volume giving a value of 49 Å$^3$, 67 Å$^3$, and 85 Å$^3$ for Ala, Abu, and Val, respectively (see Figure 7d and Supporting Information). A plot of $\Delta G^\ddagger$ vs CPK volume shows a linear relation (Figure 7d), therefore providing a good representation of the overall steric bulk. This analysis also provides a physical meaning by comparing the relative size of the cavity of the macrocycle and the CPK volume of the speed bump, which can be readily estimated in analogy to Rebek's 55% optimal occupancy rule.[42] The Ala CPK volume is smaller than the macrocycle cavity (speed bump factor = $V_{Ala}/V_{Cavity}$ = 0.80), and therefore the steric speed bump effect will be relatively small or negligible. The Abu CPK volume is similar to that of the macrocycle cavity (speed bump factor = $V_{Abu}/V_{Cavity}$ = 1.10) showing a medium steric speed bump effect, and the Val CPK volume is significantly larger than the macrocycle cavity (speed bump factor = $V_{Val}/V_{Cavity}$ = 1.39) showing larger speed bump effect (Figure 7d). Overall, the relative speed bump CPK volume with regard to the macrocycle cavity, defined as *speed bump factor* = $V_{Val}/V_{Cavity}$, analysis described here represents the observed experimental speed bump effect. We envisage that the use of speed bump factors can be a useful tool in estimating custom macrocycle shuttling rates in designer systems. (Note: variants with significantly faster or slower ring shuttling rates than the examples measured herein are incompatible with NMR analysis).

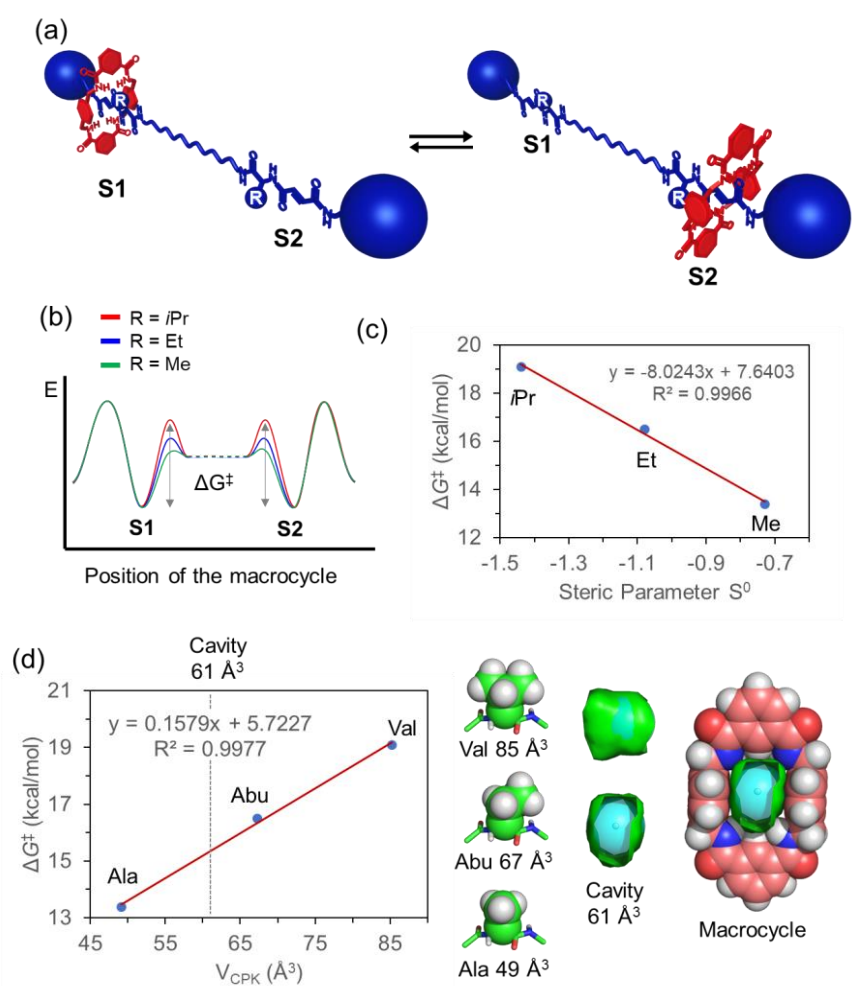

**Figure 7.** (a) Cartoon representation of the degenerate 2-station [2]rotaxanes developed in this project. (b) Schematic representation of the amino-acid speed bump effect on the ring-on-thread energy profile. (c) Experimental shuttling energy barrier vs steric parameter $S^0$ (see refs [30,40] for definition of $S^0$). (d) Experimental shuttling energy barrier vs amino-acid side chain CPK volume. Atoms used to determine the amino-acid speed bump volume and representation of the macrocycle cavity. Ala = alanine, Abu = 2-aminobutanoic acid, Val = Valine. Experimental data in TCE-$d_2$ at 25 ºC.

Further evidence of the small/negligible alanine steric speed bump effect can be obtained by comparing the 13.4 kcal/mol free energy of activation in rotaxane **1c** with more weakly binding stations - typically 11–12 kcal/mol,[12,20–22,32,43] and 16.2 kcal/mol for the high affinity fumaramide station (Figure 1).[21] The activation energy found for rotaxane **1c** is already limited by bond breaking between macrocycle and initial station and chain folding. Therefore, the size of the alanine speed bump is effectively not significantly hindering the shuttling rate.

Having measured processes in a non-competitive solvent, we further evaluated the effect of polar and protic solvents in the macrocycle shuttling rate in rotaxane **1a**. While addition of 5% MeOD to TCE-$d_2$ did not produce any measurable acceleration, the more polar mixture CDCl$_3$/MeOD/D$_2$O 45:45:10 v/v produced an increase of the shuttling rate from ca. 0.05 s$^{-1}$ to 0.14 s$^{-1}$. Then we hydrolysed the rotaxane **1a** to obtain the hexa-anionic rotaxane **1a-WS**. Unfortunately, attempts to dissolve **1a-WS** in pure D$_2$O at the concentrations required for $^1$H NMR spectroscopy were unsuccessful. Full solubilisation was achieved in MeOD/D$_2$O solutions when high proportions of MeOD were employed, i.e. MeOD/D$_2$O 9:1, v/v. The determined shuttling rate in this solvent mixture was 0.97 s$^{-1}$, highlighting the lesser effect of polar solvents on the shuttling rate in these systems, compared with steric influences (Table 2).

**Table 2.** Shuttling constants ($k_{ex}$) and the corresponding $\Delta G^{\ddagger}$ values for the rotaxanes **1a**, **1b**, **1c** and **1a-WS** in different polar solvent mixtures at 25 °C, 600 MHz

| Solvent | rotaxane | $k_{ex}$ (s$^{-1}$) | $\Delta G^{\ddagger}$ (kcal/mol) |
|---|---|---|---|
| TCE-$d_2$ | **1a**[a] | 0.07 | 19.1 |
| TCE-$d_2$ | **1b**[a] | 4.6 | 16.5 |
| TCE-$d_2$ | **1c**[a] | 980 | 13.4 |
| TCE-$d_2$/MeOD 95:5 | **1a**[b] | <0.05 | >19.3 |
| CDCl$_3$/MeOD/D$_2$O 45:45:10 | **1a**[b] | 0.14 | 18.6 |
| MeOD/D$_2$O 9:1 | **1a-WS**[b] | 0.97 | 17.5 |

[a] Determined by lineshape fitting VT $^1$H NMR and extrapolated at 25 °C. [b] Determined by EXSY $^1$H NMR.

## Conclusion

A family of degenerate 2-station [2]rotaxanes incorporating amino-acid units in conjunction with fumaramide stations is reported. In these systems the appendages on the amino-acid units act as speed bumps, which serve to slow down (up to 4 orders of magnitude) ring shuttling in accordance with a linear dependence on the appendage steric factor or CPK volume. In this regard a degree of predictability in shuttling rates is possible, profiting from a range of commercially-available amino acids with appendages presenting different steric factors. Equally noteworthy, the effect of aqueous solvents on ring shuttling is relatively small (<10) for the range of different solvent mixtures assayed, compared to the predominant steric influence.

# Experimental Section

**Materials and methods**

All chemicals and solvents were obtained from commercial sources and used without further purification unless specified. Mass spectra and NMR were performed in the CESAMO analytical facilities (Bordeaux, France). **NMR**: $^1$H and $^{13}$C NMR spectra were recorded in a Bruker Avance 300 or Bruker Avance III 600 spectrometer. Chemical shifts are reported in ppm and referenced to the solvent residual peaks. **Mass spectrometry**: Field desorption spectra (FD) were recorded on a AccuTOF (JEOL) mass spectrometer using an FD emitter with an emitter voltage of 10 kV. The sample (1–2 µL) was deposited on a 13 µm emitter wire. Electrospray spectra (ESI) were recorded on a Qexactive (Thermo) mass spectrometer. The instrument is equipped with an ESI source and spectra were recorded in positive mode. The spray voltage was maintained at 3200 V and capillary temperature set to 320 °C. Samples were introduced by injection through a 20 µL loop into a 300 µL/min flow of methanol from the LC pump.

**Synthetic procedures**

Rotaxanes were prepared using the synthetic scheme described in Figure 2. The TRIS based stopper **6** was prepared using a literature procedure.[34] Compounds **3a**, **3b**, and **3c** were prepared using literature procedures.[35,36] Signal labels used in the assignment and $^1$H and $^{13}$C NMR spectra are provided in the Supporting Information.

*Synthesis of rotaxane **1a***: A solution of *p*-xylylenediamine (35 mg, 0.255 mmol, 4 eq.) in chloroform (10 mL) and a solution of isophthaloyl dichloride (52 mg, 0.255 mmol, 4 eq.) in chloroform (10 mL) were simultaneously added at room temperature with a syringe pump during 2h to a solution of thread **2a** (100 mg, 0.0637 mmol, 1 eq.) and dry triethylamine (71 µL, 0.510 mmol, 8 eq.) in chloroform (20 mL). The reaction mixture was stirred at room temperature overnight under $N_2$ atmosphere. The reaction mixture was filtered through a Celite® pad. The filtrate was washed successively with a sodium bicarbonate solution (10%), water and brine. The organic layer was dried over magnesium sulfate, filtered and concentrated *in vacuo*. Purification by silica gel column chromatography (cyclohexane/*i*-PrOH, 90:10 to 84:16, v/v) afforded **1a** as a colourless solid (13 mg, 10% yield). **$^1$H NMR** (600 MHz, TCE-$d_2$) δ 8.55 (s, 2H, $H_u$), 8.20 (t, *J* = 7.6 Hz, 4H, $H_s$), 7.68 (m, 2H, $H_v$ or $H_{v'}$), 7.63 (t, 2H, *J* = 7.8 Hz, $H_t$), 7.61 (m, 2H, $H_v$ or $H_{v'}$), 7.10 (psq, *J* = 7.9 Hz, 8H, $H_x$), 6.94 (m, 1H, $H_h$ or $H_{h'}$), 6.88 (d, *J* = 15.0 Hz, 1H, $H_g$ or $H_{f}$), 6.87 (s, 1H, $H_{e'}$), 6.80 (d, *J* = 15.1 Hz, 1H, $H_g$ or $H_{f}$), 6.56 (d, *J* = 8.6 Hz, 1H, $H_h$ or $H_{h'}$), 6.49 (s, 1H, $H_e$), 6.06 (m, 1H, $H_l$), 5.84 (d, *J* = 14.6 Hz, 1H, $H_f$ or $H_{g'}$), 5.69 (brs, 1H, $H_{l'}$), 5.58 (d, *J* = 14.6 Hz, 1H, $H_f$ or $H_{g'}$), 4.60 (m, 4H, $H_w$), 4.40 – 4.18 (m, 5H, $H_w$ + $H_i$), 3.71 (m, 7H, $H_{i'}$ + $H_d$), 3.68 (m, 6H, $H_{d'}$), 3.64 (t, *J* = 6.3 Hz, 12H, $H_c$), 3.41 – 3.03 (m, 4H, $H_m$), 2.45 (m, 12H, $H_b$), 2.11 – 2.06 (m, 1H, $H_j$), 1.88 – 1.81 (m, 1H, $H_{j'}$), 1.50 – 1.44 (m, 4H, $H_n$), 1.43 (s, 27H, $H_a$), 1.39 (s, 27H, $H_{a'}$), 1.33 – 1.19 (m, 16H, $H_{o-r}$), 0.97 – 0.74 (m, 12H, $H_k$ + $H_{k'}$). **$^{13}$C NMR** (151 MHz, TCE-$d_2$) δ 170.95, 170.83, 170.17, 169.11, 166.03, 164.90, 164.69, 163.97, 163.71, 136.85, 136.75, 134.47, 133.54, 133.42, 131.77, 131.55, 131.30, 129.29, 129.25, 129.18, 124.33, 80.64, 80.46, 68.71, 68.60, 67.07, 66.95, 61.21, 60.09, 59.36, 58.57, 44.09, 39.71, 39.52, 36.03, 35.99, 31.11, 29.61, 29.33, 29.27, 29.03, 28.96, 27.99, 27.92, 19.14, 18.03 (1 signal overlapping/missing). **HRMS** (ESI): *m/z* calcd for $C_{112}H_{168}N_{10}O_{28}$+Na$^+$: 2124.19218 [*M*+Na]$^+$; found: 2124.19153.

*Synthesis of rotaxane **1b***: A solution of *p*-xylylenediamine (35 mg, 0.259 mmol, 4 eq.) in chloroform (10 mL) and a solution of isophthaloyl dichloride (53 mg, 0.259 mmol, 4 eq.) in chloroform (10 mL) were simultaneously added at room temperature with a syringe pump during 2h to a solution of thread **2b** (100 mg, 0.0649 mmol, 1 eq.) and dry triethylamine (72 µL, 0.519 mmol, 8 eq.) in chloroform (20 mL). The reaction mixture was stirred at room temperature overnight under $N_2$ atmosphere. The reaction mixture was filtered through a Celite® pad. The filtrate was washed successively with a sodium bicarbonate solution (10%), water and brine. The organic layer was dried over magnesium sulfate, filtered and concentrated *in vacuo*. Purification by silica gel column chromatography (cyclohexane/*i*-PrOH, 9:1 to 8:2, v/v) afforded **1b** as a colourless solid (21 mg, 16% yield). **$^1$H NMR** (600 MHz, TCE-$d_2$) δ 8.54 (s, 2H, $H_u$), 8.18 (t, $J$ = 8.0 Hz, 4H, $H_s$), 7.72 (m, 2H, $H_v$ or $H_{v'}$), 7.67 (m, 2H, $H_v$ or $H_{v'}$), 7.62 (t, $J$ = 7.8 Hz, 2H, $H_t$), 7.21 (m, 2H, $H_{h'}$), 7.11 (s, 8H, $H_x$), 6.88 (d, $J$ = 14.7 Hz, 1H, $H_f$ or $H_g$), 6.87 (s, 1H, $H_{e'}$), 6.79 (d, $J$ = 14.9 Hz, 1H, $H_f$ or $H_g$), 6.72 (d, $J$ = 7.8 Hz, 1H, $H_h$), 6.52 (s, 1H, $H_e$), 6.25 (m, 1H, $H_l$ or $H_{l'}$), 5.84 (d, $J$ = 14.9 Hz, 1H, $H_{f'}$ or $H_{g'}$), 5.82 (m, 1H, $H_l$ or $H_{l'}$), 5.61 (d, $J$ = 14.6 Hz, 1H, $H_{f'}$ or $H_{g'}$), 4.61 (m, 4H, $H_w$), 4.35 (m, 1H, $H_i$), 4.30 (m, 4H, $H_w$), 3.87 (m, 1H, $H_{i'}$), 3.71 (s, 6H, $H_d$), 3.69 (m, 6H, $H_{d'}$), 3.64 (m, 12H, $H_c$), 3.34 – 3.03 (m, 4H, $H_m$), 2.45 (m, 12H, $H_b$), 1.92 – 1.61 (dm, 4H, $H_j$), 1.50 – 1.45 (m, 4H, $H_n$), 1.43 (s, 27H, $H_a$ or $H_{a'}$), 1.39 (s, 27H, $H_a$ or $H_{a'}$), 1.33 – 1.16 (m, 16H, $H_{o-r}$), 0.93 – 0.78 (m, 6H, $H_k$ + $H_{k'}$). **$^{13}$C NMR** (151 MHz, CDCl$_3$) δ 171.13, 171.10, 171.04, 170.22, 166.71, 166.62, 165.30, 165.11, 164.45, 164.18, 137.06, 134.81, 133.88, 133.80, 132.05, 131.74, 131.51, 129.53, 129.49, 129.28, 124.78, 80.92, 80.73, 69.10, 68.96, 67.34, 67.24, 61.44, 60.48, 55.55, 54.79, 44.53, 44.47, 39.91, 39.75, 36.27, 36.26, 31.58, 30.33, 29.84, 29.57, 29.43, 28.26, 28.20, 26.04, 25.71, 10.15, 10.00 (1 signal overlapping/missing). **HRMS** (ESI): *m/z* calcd for $C_{110}H_{164}N_{10}O_{28}$+Na$^+$: 2096.160888 [*M*+Na]$^+$; found: 2096.160868.

*Synthesis of rotaxane* **1c**: A solution of *p*-xylylenediamine (36 mg, 0.264 mmol, 4 eq.) in chloroform (10 mL) and a solution of isophthaloyl dichloride (54 mg, 0.264 mmol, 4 eq.) in chloroform (10 mL) were simultaneously added at room temperature with a syringe pump during 2h to a solution of thread **2c** (100 mg, 0.0661 mmol, 1 eq.) and dry triethylamine (74 µL, 0.528 mmol, 8 eq.) in chloroform (20 mL). The reaction mixture was stirred at room temperature overnight under $N_2$ atmosphere. The reaction mixture was filtered through a Celite® pad. The filtrate was washed successively with a sodium bicarbonate solution (10%), water and brine. The organic layer was dried over magnesium sulfate, filtered and concentrated *in vacuo*. Purification by silica gel column chromatography (cyclohexane/*i*-PrOH, 9:1 to 8:2, v/v) afforded **1c** as a colourless solid (16 mg, 12% yield). **$^1$H NMR** (600 MHz, 85 °C, TCE-$d_2$) δ 8.51 (s, 2H, $H_t$), 8.24 (m, 4H, $H_r$), 7.64 (m, 2H, $H_s$), 7.57 (m, 4H, $H_u$), 7.18, (s, 8H, $H_w$), 6.53 (m, 2H, $H_h$), 6.35 (m, 2H, $H_e$), 6.28 (m, 4H, $H_f$ + $H_g$), 5.84 (m, 2H, $H_k$), 4.53 (m, 8H, $H_v$), 4.23 (m, 2H, $H_i$), 3.74 (m, 12H, $H_d$), 3.70 (m, 12H, $H_c$), 3.32 – 3.11 (m, 4H, $H_l$), 2.48 (m, 12H, $H_b$), 1.71 – 1.49 (m, 10H, $H_j$ + $H_m$), 1.47 (s, 54H, $H_a$), 1.41 – 1.20 (m, 16H, $H_{n-q}$). **$^{13}$C NMR** (151 MHz, TCE-$d_2$) δ 170.91, 166.16, 164.07, 136.88, 133.58, 131.35, 129.19, 124.34, 80.55, 68.64, 67.00, 60.57, 49.09, 44.08, 39.62, 36.00, 31.32, 30.07, 29.61, 29.21, 29.17, 28.93, 27.95 (5 signals overlapping/missing). **HRMS** (ESI): *m/z* calcd for $C_{108}H_{160}N_{10}O_{28}$+Na$^+$: 2068.12957 [*M*+Na]$^+$; found: 2068.12963.

*Synthesis of water soluble rotaxane* **1a-WS**: The valine-containing rotaxane **1a** (14 mg, 6.66 µmol) was dissolved in formic acid (1 mL) and stirred at room temperature for 6h. Volatiles were removed *in vacuo*. Methanol (0.9 mL) and a solution of sodium bicarbonate (3.4 mg, 40.0 µmol, 6 eq.) in water (0.1 mL) were added successively to the residue. Rotaxane **1a-WS** was obtained upon removal of the volatiles *in vacuo* as a colourless solid (quantitative yield). **$^1$H NMR** (600 MHz, MeOD/D$_2$O 9:1, v/v) δ 8.75 (s, 2H, $H_q$), 8.14 (dm, 4H, $H_o$), 7.70 (t, $J$ = 7.8 Hz, 2H, $H_p$), 7.16 (s, 8H, $H_s$), 6.97 (d, $J$ = 15.4 Hz, 1H, $H_d$ or $H_e$), 6.91 (d, $J$ = 15.4 Hz, 1H, $H_d$ or $H_e$), 5.97 (d, $J$ = 15.0 Hz, 1H, $H_{d'}$ or $H_{e'}$), 5.83 (d, $J$ = 15.0 Hz, 1H, $H_{d'}$ or $H_{e'}$), 4.53 – 4.36 (m, 8H, $H_r$), 4.19 (d, $J$ = 7.3 Hz, 1H, $H_f$), 3.86 (d, $J$ = 8.7 Hz, 1H, $H_{f'}$), 3.77 (s, 6H, $H_c$), 3.69 (t, $J$ = 6.8 Hz, 6H, $H_b$), 3.67 – 3.60 (m, 12H, $H_{c'}$ + $H_{b'}$), 3.25 – 3.00 (m, 4H, $H_i$), 2.42 (t, $J$ = 6.9 Hz, 12H, $H_a$), 2.13 (m, 1H, $H_g$), 1.95 (m, 1H, $H_{g'}$), 1.49 (m, 4H, $H_j$), 1.44 (m, 16H, $H_{k-n}$), 1.04 – 0.90 (m, 12H, $H_h$ + $H_{h'}$). **$^{13}$C NMR** (151 MHz, MeOD/D$_2$O 9:1, v/v) δ 180.37, 180.04, 173.40, 173.06, 169.14, 169.10, 167.25, 167.18, 167.06, 166.91, 161.35, 137.63, 137.56, 135.42, 134.94, 134.90, 132.80, 132.56, 132.46, 132.43, 130.59, 130.56, 130.36, 129.86, 126.95, 70.25, 70.15, 70.12, 70.07, 62.69, 62.02, 60.98, 45.16, 40.43, 39.38, 39.29, 30.50, 30.15, 19.68, 19.63 (3 signals overlapping/missing). **HRMS** (ESI): *m/z* calcd for $(C_{88}H_{114}N_{10}O_{28})^{6-}$+4(H$^+$)+2(Na$^+$): 905.40290 [*M*+4H+2Na]$^{2+}$; found: 905.40192.

*Synthesis of thread 2a*: To a solution of valine **3a** (199 mg, 0.5 mmol, 1 eq.), fumaric acid stopper **4** (604 mg, 1 mmol, 2 eq.), DMAP (147 mg, 1.2 mmol, 2.4 eq.) and triethylamine (0.17 mL, 1.2 mmol, 2.4 eq.) in dichloromethane (16 mL), EDCI·HCl (230 mg, 1.2 mmol, 2.4 eq.) was added in one portion at room temperature. The reaction mixture was stirred at room temperature for 16h under $N_2$ atmosphere. The reaction mixture was diluted with dichloromethane and washed successively with a citric acid solution (10%), a sodium bicarbonate solution (10%), water and brine. The organic layer was dried over magnesium sulfate, filtered and concentrated *in vacuo*. Purification by silica gel column chromatography (DCM/MeOH, 100:0 to 96:4, v/v) afforded **2a** as a colourless sticky solid (336 mg, 43% yield). **$^1$H NMR** (300 MHz, CDCl$_3$) δ 6.96 (d, $J$ = 15.0 Hz, 2H, $H_f$ or $H_g$), 6.88 (d, $J$ = 15.0 Hz, 2H, $H_f$ or $H_g$), 6.82 (d, $J$ = 8.8 Hz, 2H, $H_h$), 6.60 (s, 2H, $H_e$), 6.20 (t, $J$ = 5.7 Hz, 2H, $H_l$), 4.25 (pst, $J$ = 8.9 Hz, 2H, $H_i$), 3.74 (s, 12H, $H_d$), 3.64 (t, $J$ = 6.3 Hz, 12H, $H_c$), 3.23 (m, 4H, $H_m$), 2.44 (t, $J$ = 6.1 Hz, 12H, $H_b$), 2.13 (m, 2H, $H_j$), 1.53 – 1.46 (m, 4H, $H_n$), 1.44 (s, 54H, $H_a$), 1.34 – 1.19 (m, 16H, $H_{o-r}$), 0.95 (d, $J$ = 6.8 Hz, 6H, $H_k$ or $H_{k'}$), 0.93 (d, $J$ = 6.8 Hz, 6H, $H_k$ or $H_{k'}$). **$^{13}$C NMR** (151 MHz, CDCl$_3$) δ 171.04, 170.82, 164.62, 164.26, 134.74, 132.33, 80.65, 69.03, 67.21, 60.51, 59.30, 39.63, 36.25, 31.00, 29.53, 29.34, 29.13, 28.24, 26.87, 19.42, 18.51. **HRMS** (ESI): *m/z* calcd for C$_{80}$H$_{140}$N$_6$O$_{24}$+Na$^+$: 1591.98112 [*M*+Na]$^+$; found: 1591.98344.

*Synthesis of thread 2b*: To a solution of valine **3b** (185 mg, 0.5 mmol, 1 eq.), fumaric acid stopper **4** (604 mg, 1 mmol, 2 eq.), DMAP (147 mg, 1.2 mmol, 2.4 eq.) and triethylamine (0.17 mL, 1.2 mmol, 2.4 eq.) in dichloromethane (16 mL), EDCI·HCl (230 mg, 1.2 mmol, 2.4 eq.) was added in one portion at room temperature. The reaction mixture was stirred at room temperature for 24h under $N_2$ atmosphere. The mixture was diluted with dichloromethane and washed successively with a citric acid solution (10%), a sodium bicarbonate solution (10%), water and brine. The organic layer was dried over magnesium sulfate, filtered and concentrated *in vacuo*. Purification by silica gel column chromatography (DCM/MeOH, 100:0 to 96:4, v/v) afforded **2b** as a colourless sticky solid (372 mg, 48% yield). **$^1$H NMR** (300 MHz, CDCl$_3$) δ 6.95 (d, $J$ = 15.0 Hz, 2H, $H_f$ or $H_g$), 6.84 (d, $J$ = 15.0 Hz, 2H, $H_f$ or $H_g$), 6.76 (d, $J$ = 8.0 Hz, 2H, $H_h$), 6.61 (s, 2H, $H_e$), 6.23 (t, $J$ = 5.7 Hz, 2H, $H_l$), 4.39 (q, $J$ = 7.1 Hz, 2H, $H_i$), 3.74 (s, 12H, $H_d$), 3.65 (t, $J$ = 6.3 Hz, 12H, $H_c$), 3.23 (m, 4H, $H_m$), 2.44 (t, $J$ = 6.3 Hz, 12H, $H_b$), 1.99 – 1.61 (dm, $J$ = 71.3 Hz, 4H, $H_j$), 1.54 – 1.47 (m, 4H, $H_n$), 1.44 (s, 54H, $H_a$), 1.34 – 1.20 (m, 16H, $H_{o-r}$), 0.92 (t, $J$ = 7.4 Hz, 6H, $H_k$). **$^{13}$C NMR** (101 MHz, CDCl$_3$) δ 171.16, 171.04, 164.55, 164.23, 134.75, 132.20, 80.66, 69.09, 67.23, 60.52, 54.91, 39.67, 36.27, 29.56, 29.38, 29.17, 28.25, 26.85, 25.62, 10.09 (1 signal overlapping/missing). **HRMS** (ESI): *m/z* calcd for C$_{78}$H$_{136}$N$_6$O$_{24}$+Na$^+$: 1563.94982 [*M*+Na]$^+$; found: 1563.95145.

*Synthesis of thread 2c*: To a solution of valine **3c** (171 mg, 0.5 mmol, 1 eq.), fumaric acid stopper **4** (604 mg, 1 mmol, 2 eq.), DMAP (147 mg, 1.2 mmol, 2.4 eq.) and triethylamine (0.17 mL, 1.2 mmol, 2.4 eq.) in dichloromethane (16 mL), EDCI·HCl (230 mg, 1.2 mmol, 2.4 eq.) was added in one portion at room temperature. The reaction mixture was stirred at room temperature for 24h under $N_2$ atmosphere. The mixture was diluted with dichloromethane and washed successively with a citric acid solution (10%), a sodium bicarbonate solution (10%), water and brine. The organic layer was dried over magnesium sulfate, filtered and concentrated *in vacuo*. Purification by silica gel column chromatography (DCM/MeOH, 100:0 to 96:4, v/v) afforded **2c** as a colourless sticky solid (375 mg, 50% yield). **$^1$H NMR** (300 MHz, CDCl$_3$) δ 6.96 (d, $J$ = 15.0 Hz, 2H, $H_f$ or $H_g$), 6.87 (m, 2H, $H_h$), 6.82 (d, $J$ = 15.2 Hz, 2H, $H_f$ or $H_g$), 6.64 (s, 2H, $H_e$), 6.37 (t, $J$ = 5.7 Hz, 2H, $H_k$), 4.53 (m, 2H, $H_i$), 3.74 (s, 12H, $H_d$), 3.64 (t, $J$ = 6.3 Hz, 12H, $H_c$), 3.23 (q, $J$ = 6.7 Hz, 4H, $H_l$), 2.44 (t, $J$ = 6.2 Hz, 12H, $H_b$), 1.56 – 1.46 (m, 4H, $H_m$), 1.44 (s, 54H, $H_a$), 1.39 (d, $J$ = 6.9 Hz, 6H, $H_j$) 1.33 – 1.17 (m, 16H, $H_{n-q}$). **$^{13}$C NMR** (101 MHz, CDCl$_3$) δ 171.85, 171.07, 164.41, 164.20, 134.85, 132.05, 80.68, 69.10, 67.23, 60.51, 49.21, 39.71, 36.26, 29.52, 29.39, 29.18, 28.25, 26.83, 18.20 (1 signal overlapping/missing). **HRMS** (ESI): *m/z* calcd for C$_{76}$H$_{132}$N$_6$O$_{24}$+Na$^+$: 1535.91852 [*M*+Na]$^+$; found: 1535.92002.

*Synthesis of 4*: To a solution of methyl ester **5** (920 mg, 1.49 mmol, 1 eq.) in a mixture of THF (100 mL) and water (12 mL), a solution of LiOH·H$_2$O (118 mg, 4.47 mmol, 3 eq.) in water (5 mL) was added. The reaction mixture was stirred under reflux for 30 min and the reaction progression was monitored by TLC (cyclohexane/EtOAc 1:1). THF was evaporated *in vacuo*. The aqueous solution was acidified with a 1M HCl solution to pH ≈ 1 and extracted with dichloromethane. The organic layer was washed with brine, dried over magnesium sulfate, filtered and concentrated *in vacuo*. Purification by silica gel column chromatography (cyclohexane/EtOAc, 9:1 + 0.05% AcOH, v/v) afforded fumaric acid stopper **4** as a thick colourless oil (699 mg, 78% yield). **$^1$H NMR** (300 MHz, CDCl$_3$) δ 7.21 (d, *J* = 15.4 Hz, 1H, *H$_f$* or *H$_g$*), 7.03 (s, 1H, *H$_e$*), 6.73 (d, *J* = 15.4 Hz, 1H, *H$_f$* or *H$_g$*), 3.76 (s, 6H, *H$_d$*), 3.65 (t, *J* = 6.1 Hz, 6H, *H$_c$*), 2.45 (t, *J* = 6.1 Hz, 6H, *H$_b$*), 1.44 (s, 27H, *H$_a$*) (1 signal overlapping/missing). **$^{13}$C NMR** (151 MHz, CDCl$_3$) δ 171.35, 169.94, 163.67, 139.66, 129.04, 80.88, 69.11, 67.25, 60.58, 36.13, 28.20. **HRMS** (ESI): *m/z* calcd for C$_{29}$H$_{49}$NO$_{12}$+Na$^+$: 626.31470 [*M*+Na]$^+$; found: 626.31508.

*Synthesis of 5*: To a solution of amine stopper **6** (1.00 g, 1.98 mmol, 1 eq.), monomethyl fumarate (283 mg, 2.18 mmol, 1.1 eq.) and DMAP (293 mg, 2.40 mmol, 1.2 eq.) in dichloromethane (45 mL), EDCI·HCl (460 mg, 2.40 mmol, 1.2 eq.) was added in one portion at room temperature. The reaction mixture was stirred at room temperature for 48h under N$_2$ atmosphere and monitored by TLC (EtOAc/cyclohexane 3:1). The mixture was washed successively with a citric acid solution (10%), a sodium bicarbonate solution (10%), water and brine. The organic layer was dried over magnesium sulfate, filtered and concentrated *in vacuo*. Purification by silica gel column chromatography (cyclohexane/EtOAc, 7:3, v/v) afforded methyl ester **5** as a colourless oil (951 mg, 78% yield). **$^1$H NMR** (300 MHz, CDCl$_3$) δ 7.10 (d, *J* = 15.4 Hz, 1H, *H$_f$* or *H$_g$*), 6.88 (s, 1H, *H$_e$*), 6.74 (d, *J* = 15.4 Hz, 1H, *H$_f$* or *H$_g$*), 3.76 (s, 3H, *H$_h$*), 3.75 (s, 6H, *H$_d$*), 3.65 (t, *J* = 6.2 Hz, 6H, *H$_c$*), 2.44 (t, *J* = 6.2 Hz, 6H, *H$_b$*), 1.45 (s, 27H, *H$_a$*). **$^{13}$C NMR** (76 MHz, CDCl$_3$) δ 171.26, 166.22, 163.73, 138.19, 129.24, 80.78, 69.15, 67.26, 60.54, 52.04, 36.18, 28.24. **HRMS** (ESI): *m/z* calcd for C$_{30}$H$_{51}$NO$_{12}$+H$^+$: 618.34840 [*M*+H]$^+$; found: 618.34733.

**Dynamic NMR**

Line-broadening dynamic NMR (DNMR) fittings were performed using the DNMR module in Bruker TopSpin 4.0.6. The fumaramide and *t*Bu signals were suitable to perform the fittings to obtain the experimental exchange rate constants ($k_{ex}$).[44] The first-order exchange rate constants ($k_{ex}$) were obtained from the intensities of the EXSY diagonal peaks (I$_{AA}$ and I$_{BB}$) and the EXSY cross peaks (I$_{AB}$ and I$_{BA}$) at different mixing times ($t_{mix}$) as described in the literature to obtain a straight line with slope $k_{ex}$.[45,46] Further details and sample spectra are provided in the Supporting Information.

**Molecular Modelling**

Basic modelling was carried out to evaluate the steric size of the amino-acid speed bumps. Molecular modelling was carried out with the Spartan '18 software using the MMFF force field.[47] See Supporting information for further details.

# Acknowledgements

This project has received funding from the European Union's Horizon 2020 research and innovation programme under the Marie Skłodowska-Curie grant agreement No 796612. Equally, financial support from the Agence Nationale de la Recherche (project ANR-16-CE29-0011) and CNRS is gratefully acknowledged.

**Keywords:** rotaxanes • molecular shuttles • dynamics • EXSY NMR • hydrogen bonding

# Entry for the Table of Contents

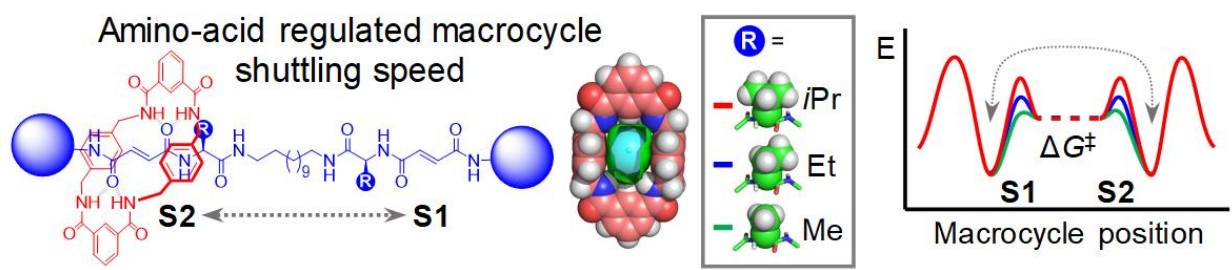

A family of degenerate 2-station [2]rotaxanes incorporating amino-acid units in conjunction with fumaramide stations is reported in aqueous and pure organic solvents. Appendages on the amino-acid units act as speed bumps, which serve to slow down (up to 4 orders of magnitude) ring shuttling in accordance with a linear dependence on the appendage steric factor and/or CPK volume.

Institute and/or researcher Twitter usernames:

@mcclenaghan_lab

@vtemarticent

# Supporting information

# Dynamic NMR experiments

Line-broadening dynamic NMR (DNMR) fittings were performed using the DNMR module in Bruker TopSpin 4.0.6. The fumaramide and *t*Bu were signals suitable to perform the fittings to obtain the experimental exchange rate constants ($k_{ex}$).[S1]

The $\Delta G^{\ddagger}$ at 25 °C was obtained to all systems by interpolation/extrapolation using Eyring plot using from the measured exchange rate constants (equations 1 and 2).[S2]

$$R \ln\left(\frac{k_{ex}}{T}\right) + R \ln\left(\frac{h}{k_B}\right) = -\Delta H^{\ddagger}\left(\frac{1}{T}\right) + \Delta S^{\ddagger} \quad Equation\ 1$$

With $R$ = 1.99 cal mol$^{-1}$ K$^{-1}$, $h$ = 6.63 × 10$^{-34}$ J s and $k_B$ = 1.23 × 10$^{-23}$ J K$^{-1}$ and a factor of 1000 in the enthalpic term to use the kcal mol$^{-1}$ units in $\Delta H^{\ddagger}$, the equation 1 converts to equation 2.

$$1.99 \ln\left(\frac{k_{ex}}{T}\right) - 47.18\ \text{cal/(mol K)} = -\Delta H^{\ddagger}\left(\frac{1000}{T}\right) + \Delta S^{\ddagger} \quad Equation\ 2$$

EXSY experiments were performed using the Bruker Topspin's standard *noesyph* pulse sequence. Different spectra were recorded using different mixing from 5 to 1000 ms. Peak 2D integrals were obtained for each spectrum to obtain the corresponding exchange rates.

The first-order exchange rate constants ($k_{ex}$) were obtained from the intensities of the EXSY diagonal peaks ($I_{AA}$ and $I_{BB}$) and the EXSY cross peaks ($I_{AB}$ and $I_{BA}$) using equations 3 and 4. Plotting ln((r+1)/(r-1)) vs $t_{mix}$ gives a straight line with slope $k_{ex}$.[S3,S4]

$$r = \frac{I_{AA} + I_{BB}}{I_{AB} + I_{BA}} \quad Equation\ 3$$

$$\ln\frac{r+1}{r-1} = k_{ex} \times t_{mix} \quad Equation\ 4$$

# VT NMR and lineshape fitting of **1a**, **1b** and **1c** in TCE-$d_2$

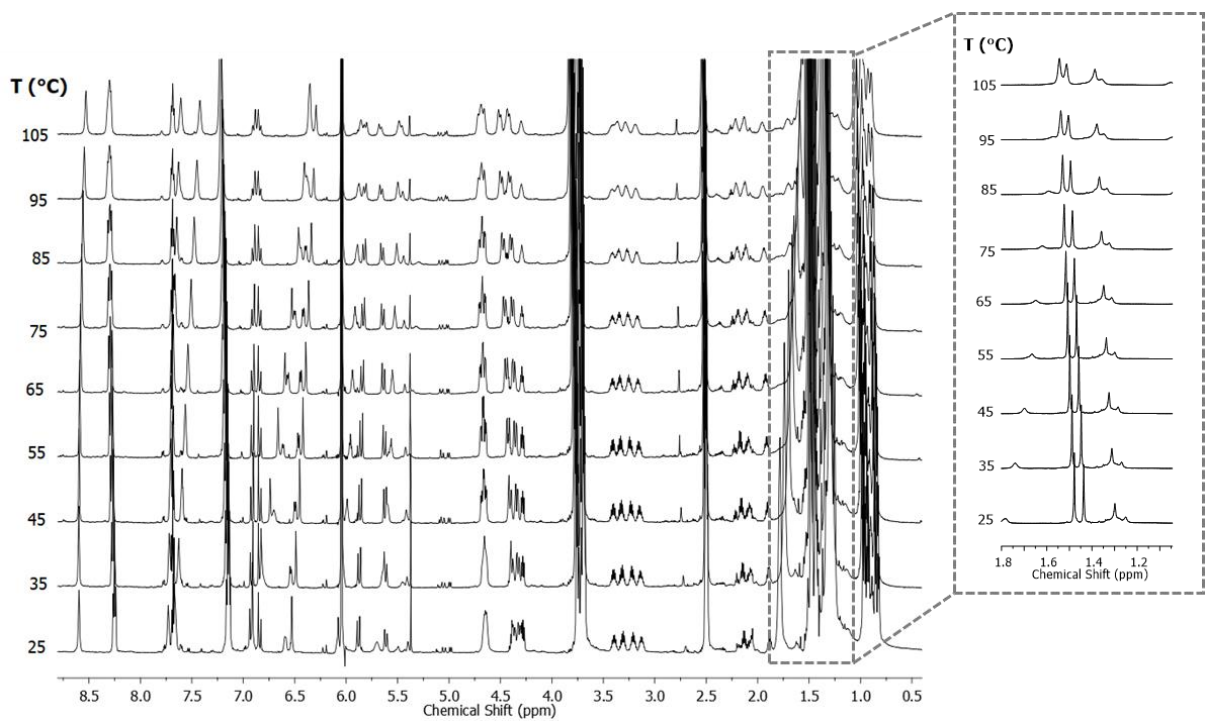

**Figure S1**. VT $^1$H-NMR of valine-containing thread **1a** (600 MHz in TCE-$d_2$). Inset shows the tBu signals.

**Table S1**. Summary of the first-order exchange rate constants ($k_{ex}$) obtained by DNMR line-shape analysis rotaxane **1a** in TCE-$d_2$.

| Temperature (°C) | $k_{ex}$ (s$^{-1}$) |
|---|---|
| 65 | 0.8 |
| 75 | 1.2 |
| 85 | 2.5 |
| 95 | 3.2 |
| 105 | 4.7 |

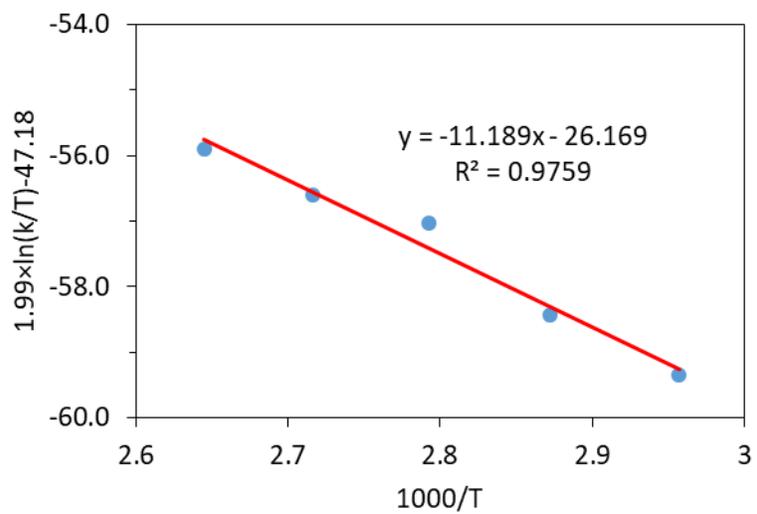

**Figure S2.** Arrhenius plot of rotaxane **1a** in TCE-$d_2$.

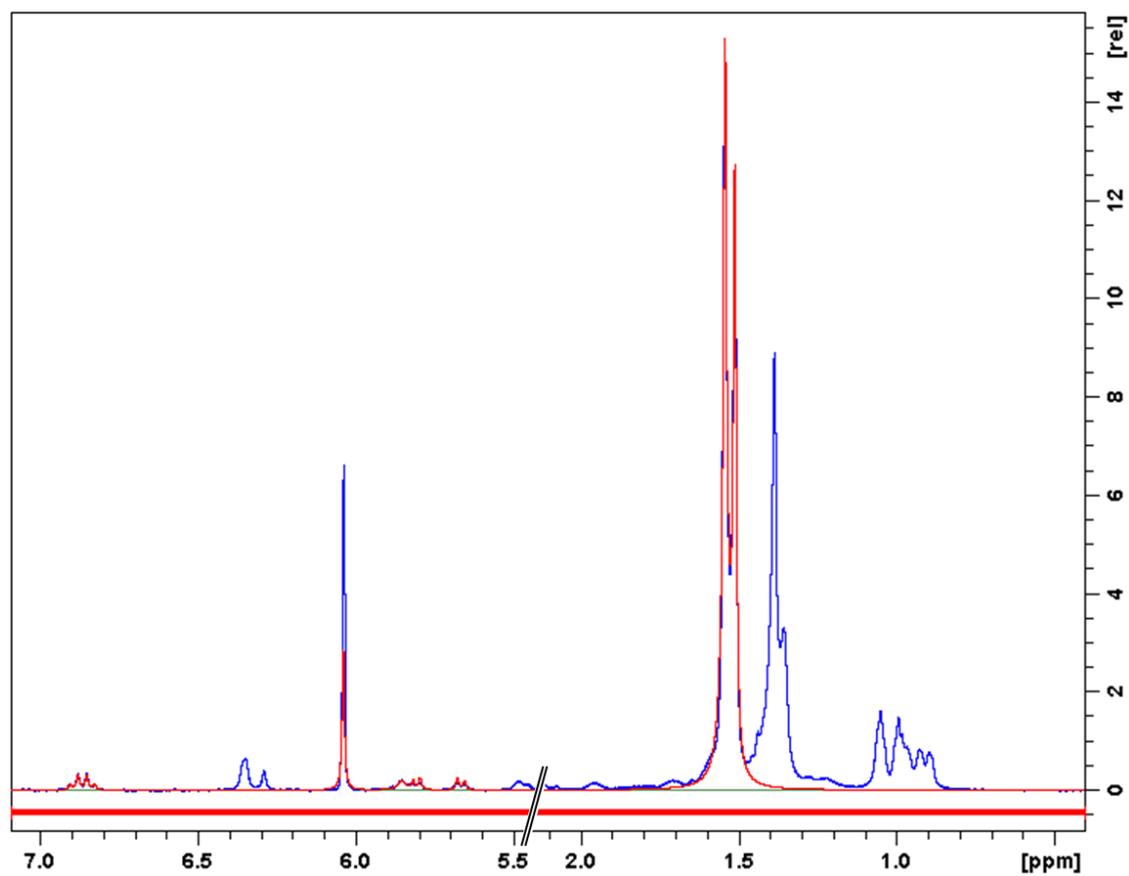

**Figure S3.** $^1$H NMR of rotaxane **1a** (blue) and DNMR fitting (red) in TCE-$d_2$ at 105 °C.

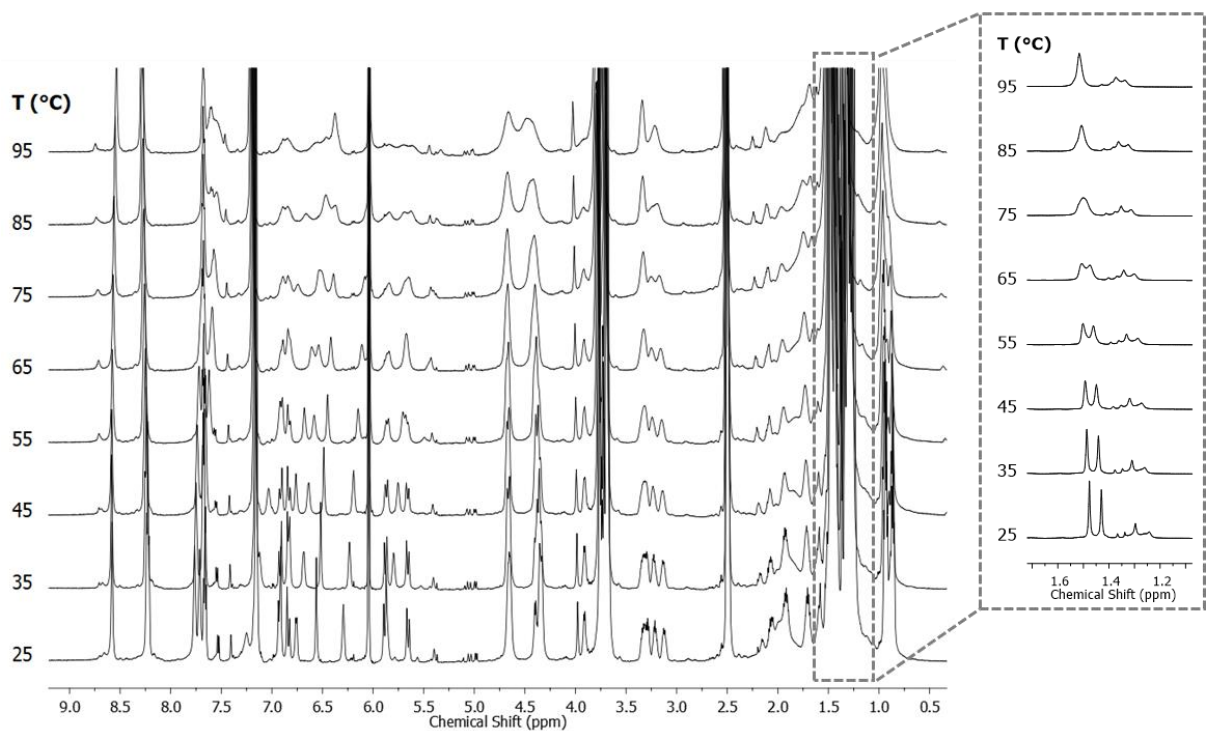

**Figure S4**. VT $^1$H-NMR of aminobutyric-containing thread **1b** (600 MHz in TCE-$d_2$). Inset shows the tBu signals.

**Table S2**. Summary of the first-order exchange rate constants ($k_{ex}$) obtained by DNMR line-shape analysis rotaxane **1b** in TCE-$d_2$.

| Temperature (°C) | $k_{ex}$ (s$^{-1}$) |
|---|---|
| 25 | 5.3 |
| 35 | 6.7 |
| 45 | 11.0 |
| 55 | 18.6 |
| 65 | 29.4 |
| 75 | 44.6 |
| 85 | 59.0 |
| 95 | 83.8 |

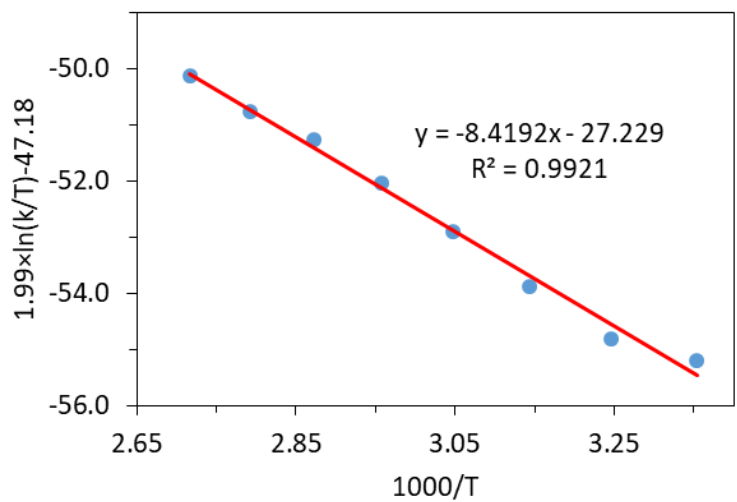

**Figure S5**. Arrhenius plot of rotaxane **1b** in TCE-$d_2$.

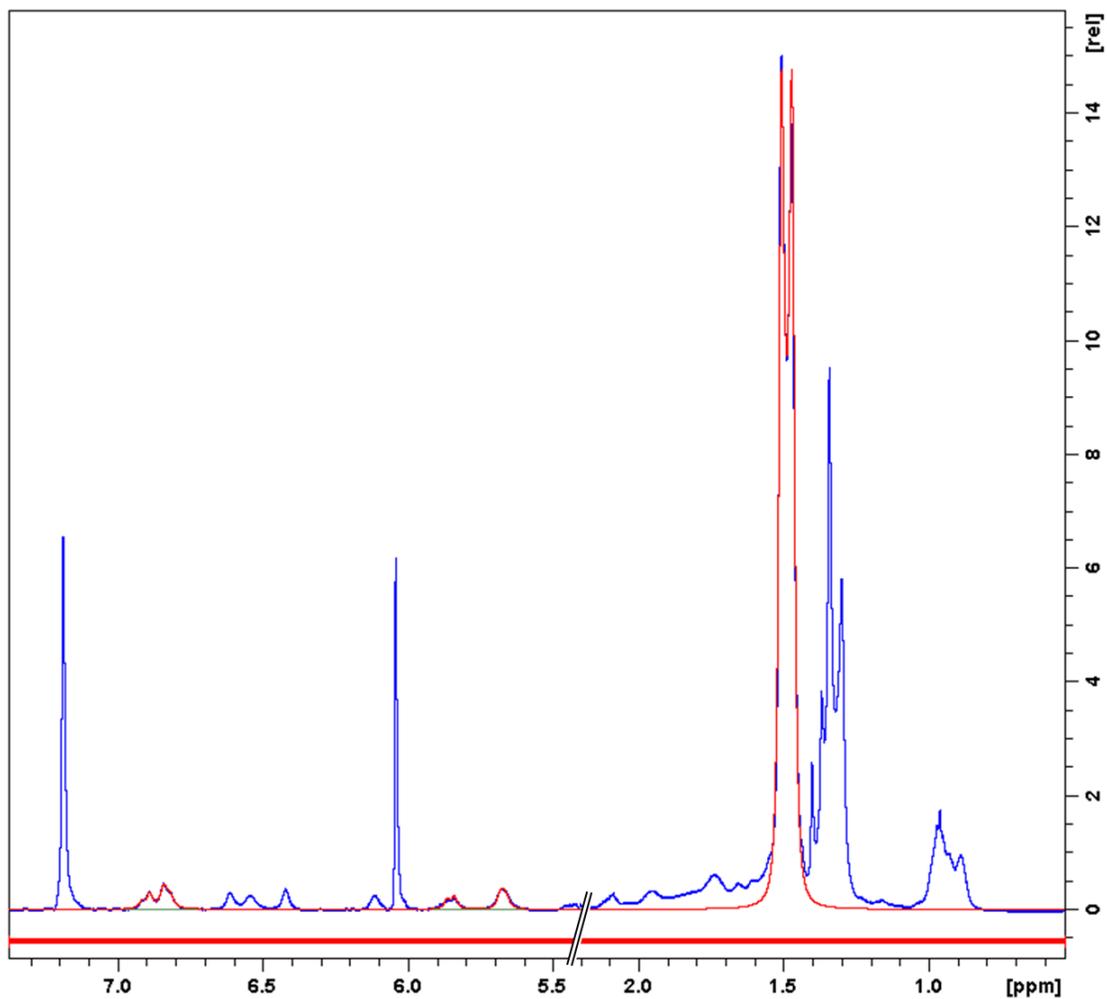

**Figure S6**. $^1$H NMR of rotaxane **1b** (blue) and DNMR fitting (red) in TCE-$d_2$ at 65 °C.

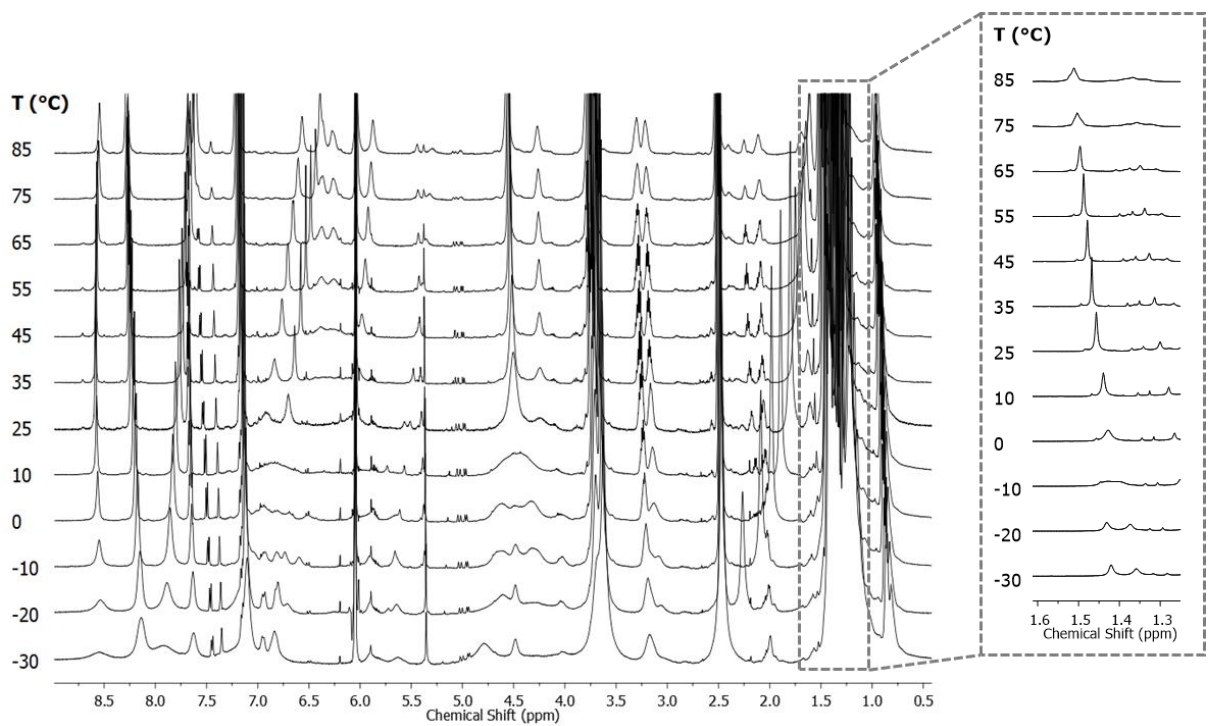

**Figure S7.** VT $^1$H-NMR of alanine-containing thread **1c** (600 MHz in TCE-$d_2$). Inset shows the tBu signals.

**Table S3.** Summary of the first-order exchange rate constants ($k_{ex}$) obtained by DNMR line-shape analysis rotaxane **1c** in TCE-$d_2$.

| Temperature (°C) | $k_{ex}$ (s$^{-1}$) |
|---|---|
| −30 | 17 |
| −20 | 20 |
| −10 | 51 |
| 0 | 161 |
| 10 | 461 |
| 25 | 844 |
| 35 | 1963 |
| 45 | 4370 |
| 55 | 6475 |
| 65 | 8472 |
| 75 | 20583 |
| 85 | 21800 |

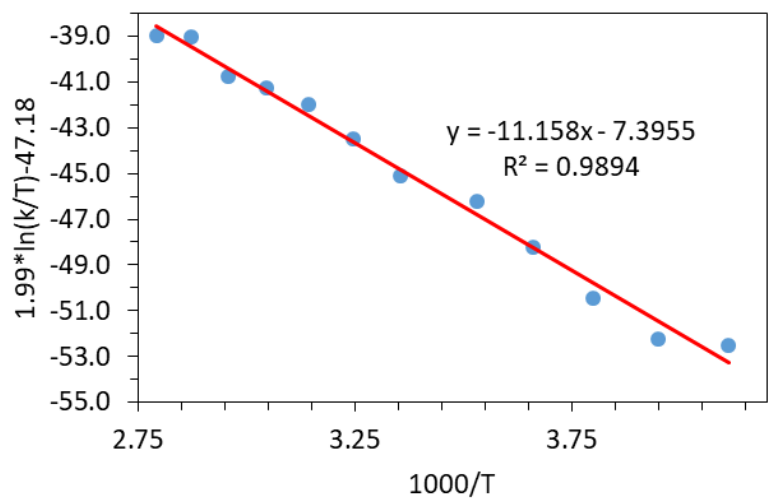

**Figure S8**. Arrhenius plot of rotaxane **1c** in TCE-$d_2$.

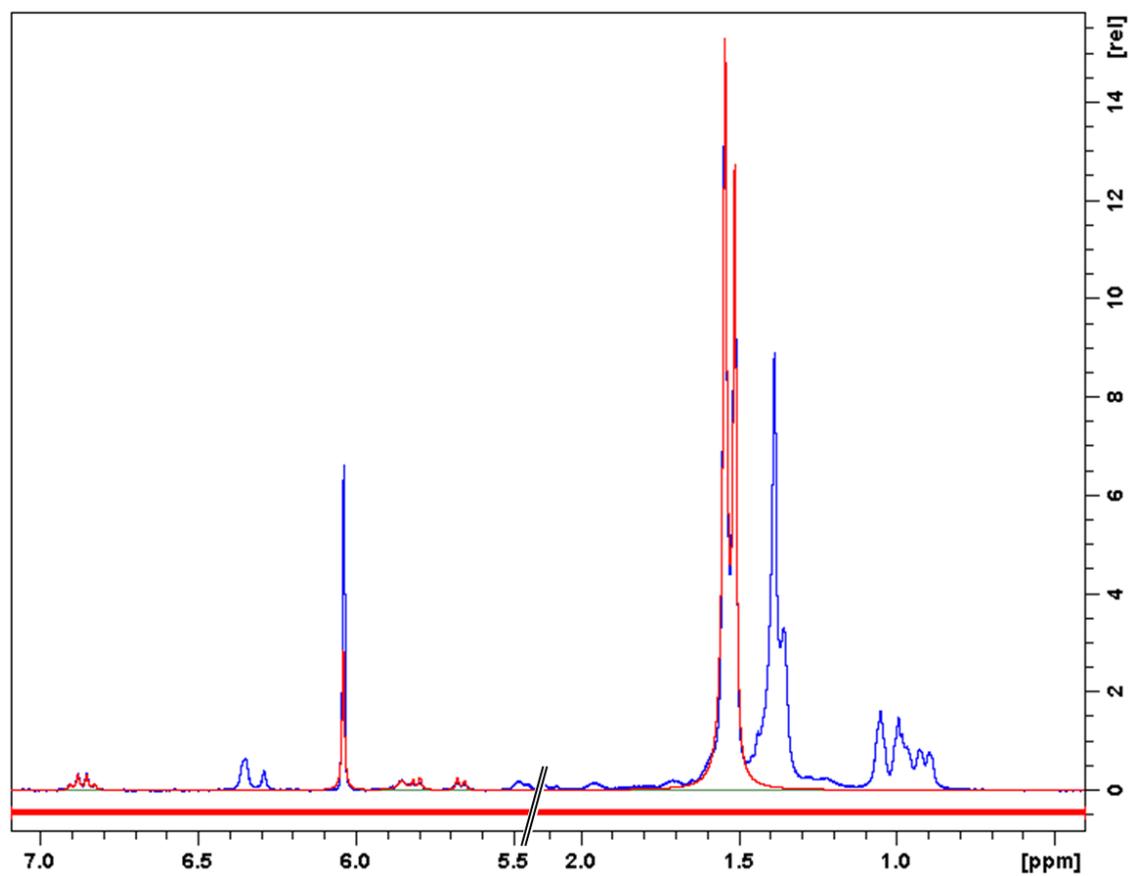

**Figure S9**. $^1$H NMR of rotaxane **1c** (blue) and DNMR fitting (red) in TCE-$d_2$ at 65 °C.

# EXSY NMR of **1a**, **1b** and **1c** in TCE-$d_2$

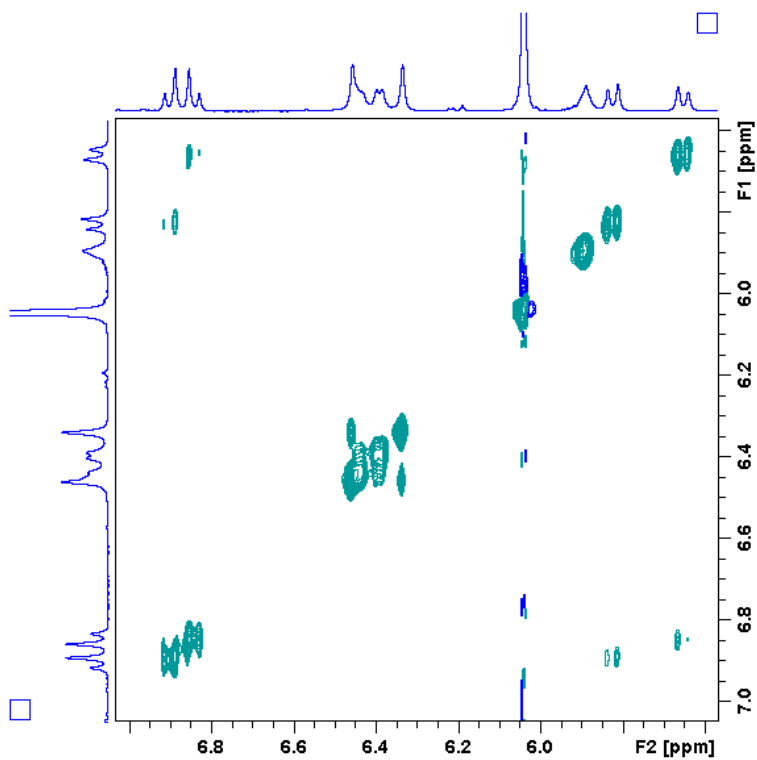

**Figure S10**. EXSY $^1$H NMR of rotaxane **1a** in TCE-$d_2$ at 85 °C, 600 MHz, 100 ms mixing time.

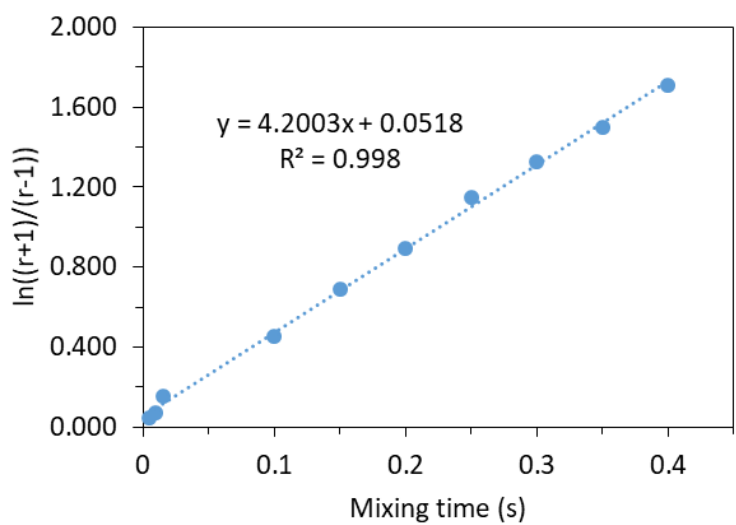

**Figure S11**. Fitting to obtain the exchange rate from EXSY $^1$H NMR of rotaxane **1a** in TCE-$d_2$ at 85 °C, 600 MHz.

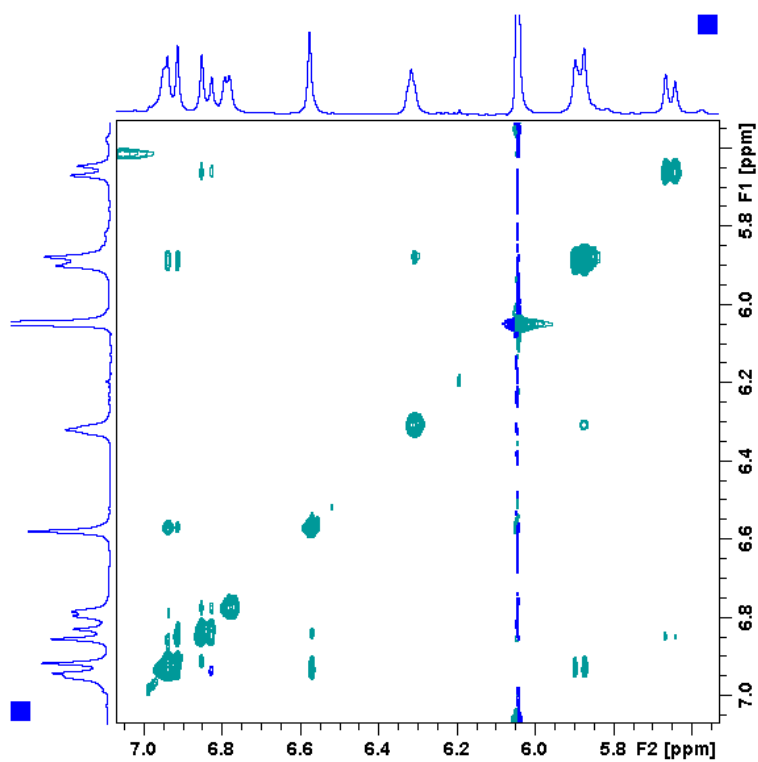

**Figure S12**. EXSY $^1$H NMR of rotaxane **1b** in TCE-$d_2$ at 25 °C, 600 MHz, 150 ms mixing time.

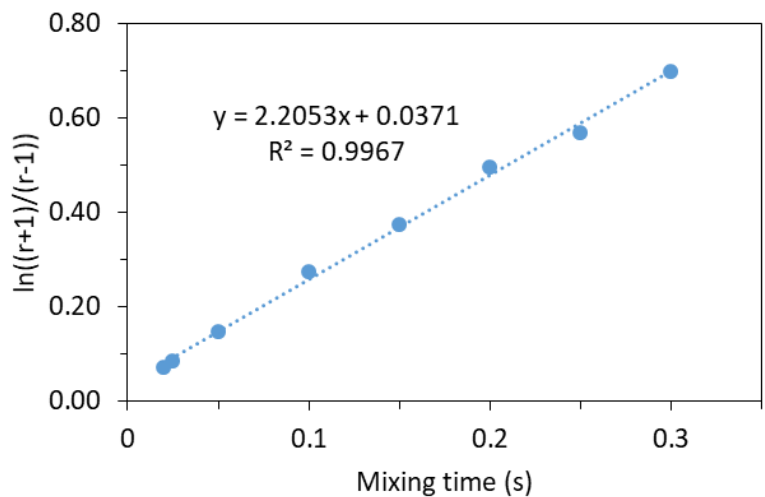

**Figure S13**. Fitting to obtain the exchange rate from EXSY $^1$H NMR of rotaxane **1b** in TCE-$d_2$ at 25 °C, 600 MHz.

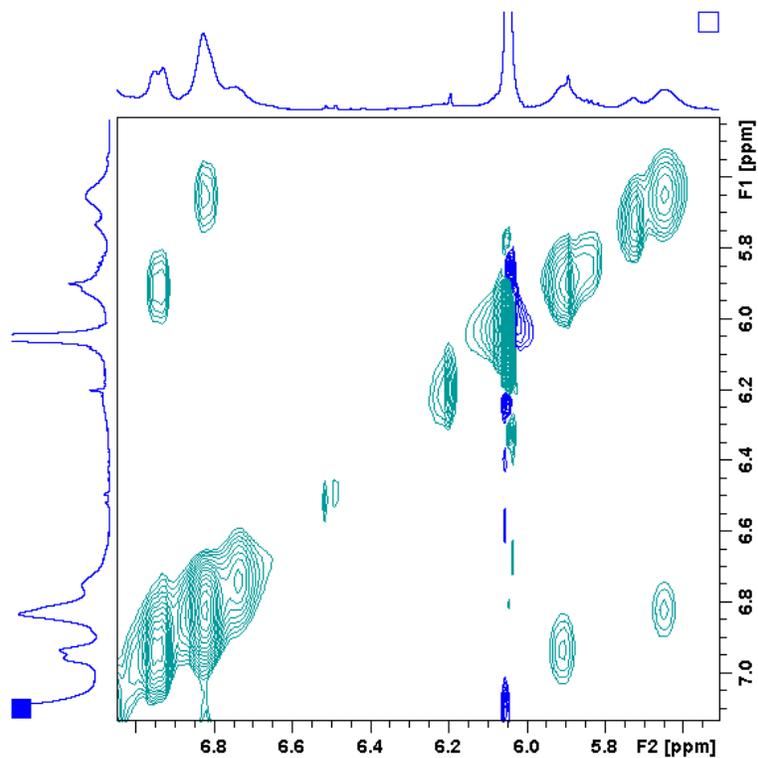

**Figure S14**. EXSY $^1$H NMR of rotaxane **1c** in TCE-$d_2$ at −20 °C, 600 MHz, 20 ms mixing time.

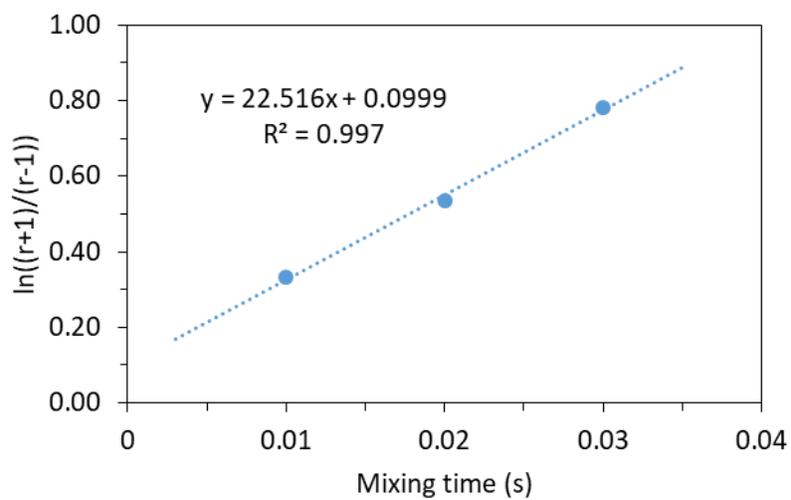

y = 22.516x + 0.0999
R² = 0.997

**Figure S15**. Fitting to obtain the exchange rate from EXSY $^1$H NMR of rotaxane **1c** in TCE-$d_2$ at −20 °C, 600 MHz.

# EXSY NMR of **1a** and **1a-WS** in polar solvent mixtures

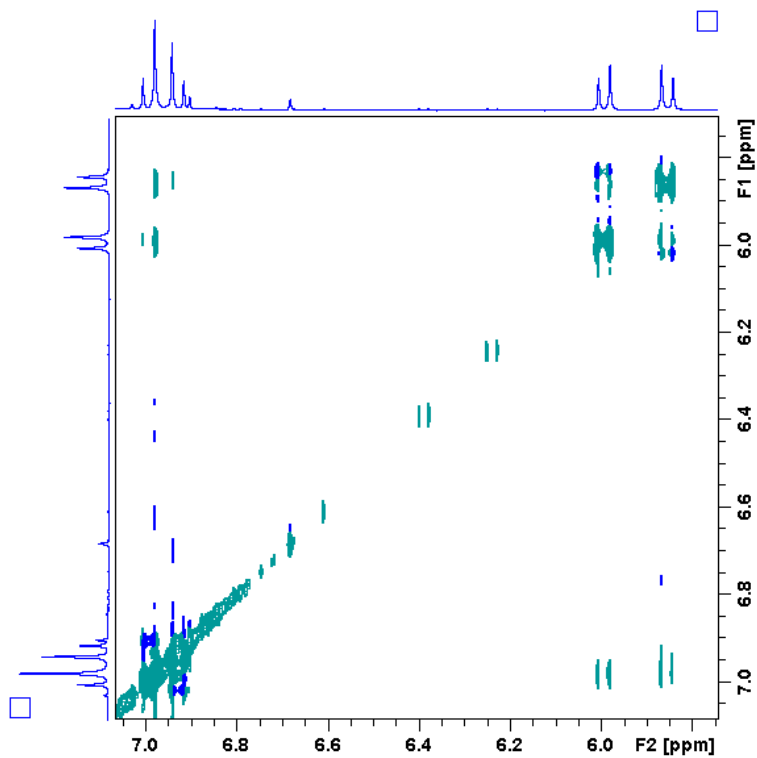

**Figure S16**. EXSY $^1$H NMR of rotaxane **1a-WS** in MeOD/D$_2$O 9:1 at 25 °C, 600 MHz, 100 ms mixing time.

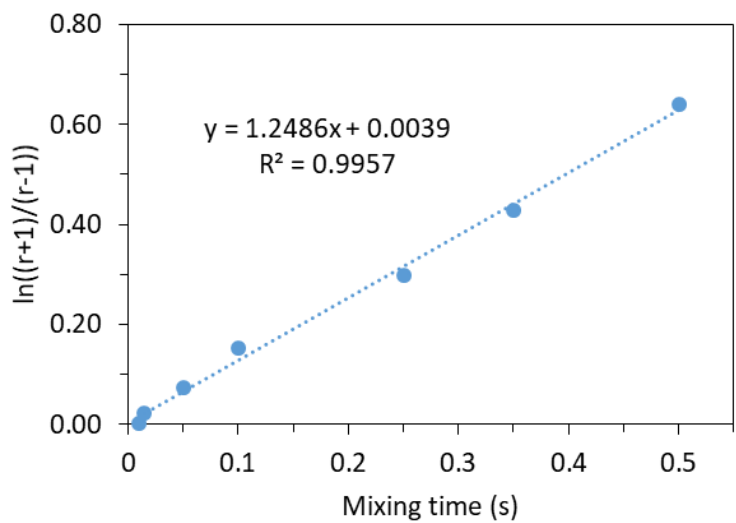

**Figure S17**. Fitting to obtain the exchange rate from EXSY $^1$H NMR of rotaxane **1a-WS** in MeOD/D$_2$O 9:1 at 25 °C, 600 MHz.

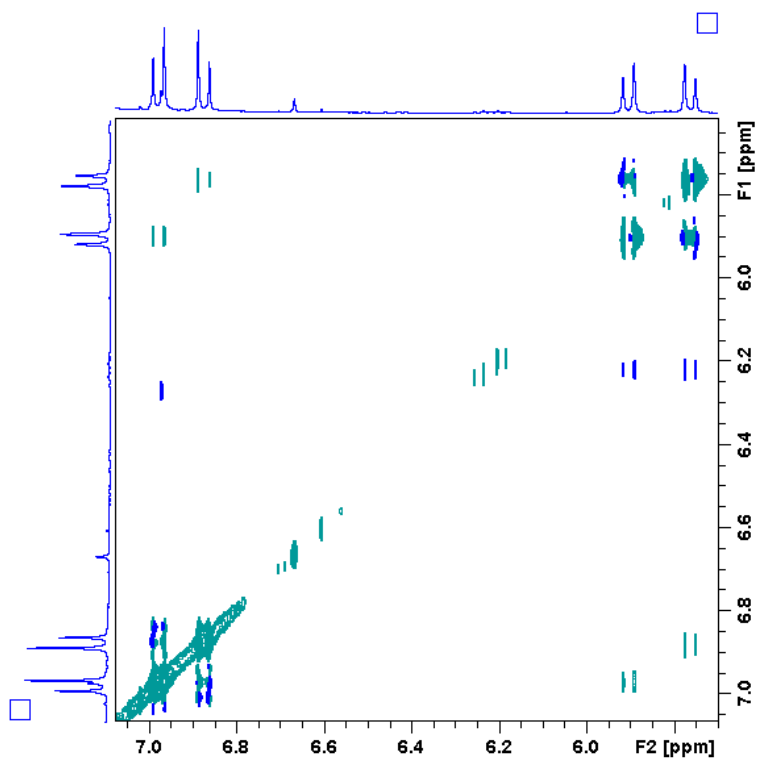

**Figure S18**. EXSY $^1$H NMR of rotaxane **1a** in CDCl$_3$/MeOD/D$_2$O 45:45:10 at 25 °C, 600 MHz, 500 ms mixing time.

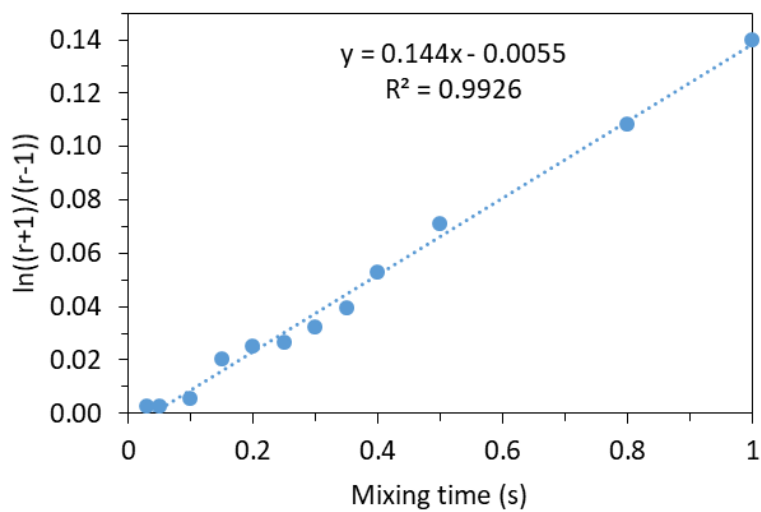

**Figure S19**. Fitting to obtain the exchange rate from EXSY $^1$H NMR of rotaxane **1a** in CDCl$_3$/MeOD/D$_2$O 45:45:10 at 25 °C, 600 MHz.

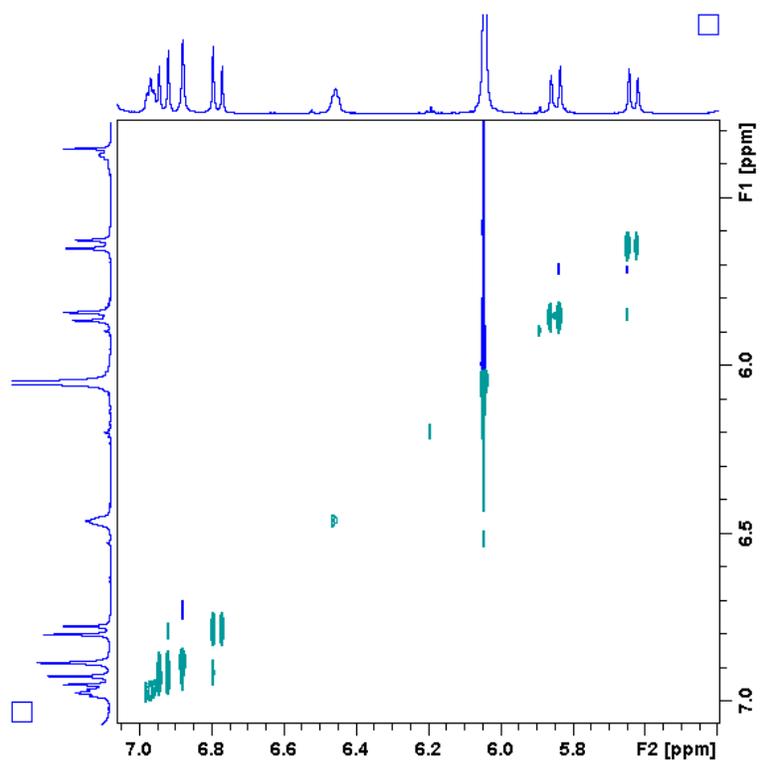

**Figure S20**. EXSY $^1$H NMR of rotaxane **1a** in TCE-$d_2$/MeOD 95:5 at 25 °C, 600 MHz, 800 ms mixing time. Exchange rate constant estimated to be <0.05 s$^{-1}$.

# Molecular Modelling

Molecular modelling was carried out with the Spartan '18 software using the MMFF force field.[S5] The isophthalamide motif in the macrocycle was constrained to be planar.

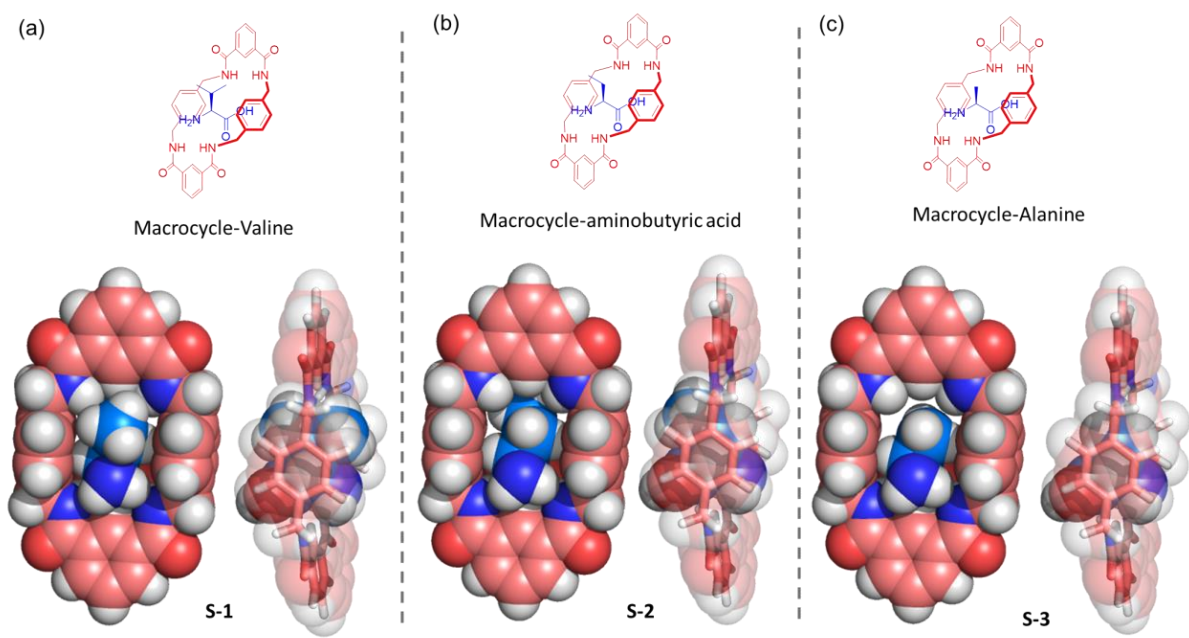

**Figure S21**. Front and side views of CPK models of pseudorotaxanes incorporating different amino acids used to evaluate their relative steric size (MMFF, Spartan '18 software): (a) Optimized structure of macrocycle-valine pseudorotaxane **S-1** with the macrocycle passing over the amino acid. (b) Structure of macrocycle-aminobutyric acid pseudorotaxane **S-2** obtained by replacement of one methyl group of **S-1** by one hydrogen atom. (c) Structure of macrocycle-Alanine pseudorotaxane **S-3** obtained by replacement of two methyl groups of **S-1** by two hydrogen atoms.

S16

# Steric parameter and CPK volumes

CPK volumes of amino-acid side chain were determined with the Spartan '18 software as described in the manuscript.[S5] The steric parameters $S^0$ were taken from the literature.[S6]

**Table S4** CPK volumes determined with the Spartan '18 software and steric parameters $S^0$ from reference [S6].

| R | Amino acid | Amino-acid side chain CPK volume $V_{CPK}$ (Å³) | Steric Parameter $S^0$ |
|---|---|---|---|
| H | Glycine | 30.54 | 0 |
| Me | Alanine | 49.12 | -0.73 |
| Et | 2-Aminobutiric acid | 67.38 | -1.08 |
| iPr | Valine | 85.35 | -1.44 |
| tBu | tert-Butylglycine | 102.98 | -3.94 |
| iBu | Leucine | 106.26 | -1.29 |
| Ph | Phenylglicine | 114.92 | -1.82 |
| CH₂Ph | Phenylalanine | 133.04 | -1.16 |

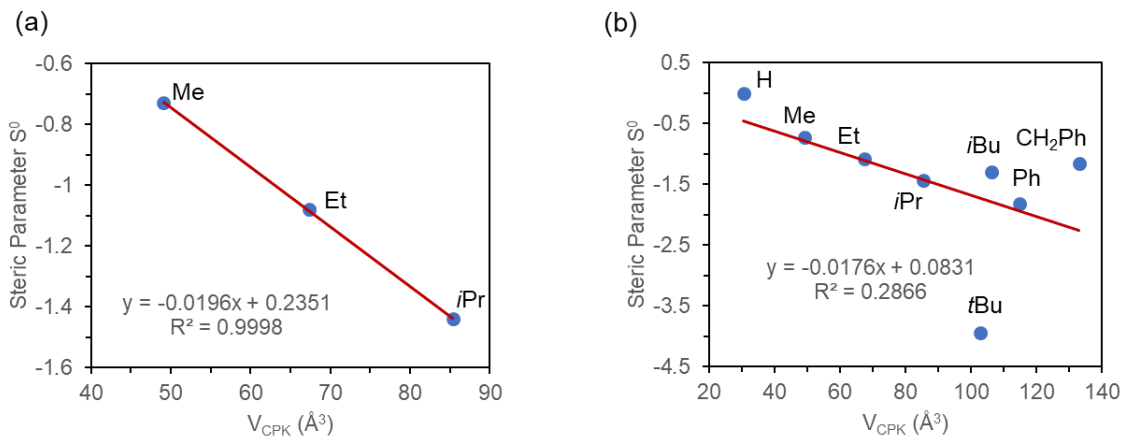

**Figure S22**. Comparison of the steric parameter $S^0$ and the CPK volume. (a) Good correlation of $S^0$ and $V_{CPK}$ for R = Me, Et, and iPr (used in this work). (b) Deviation of the substituents with larger total steric bulk (R = tBu, iBu, and CH₂Ph) to the $S^0$ and $V_{CPK}$ correlation.

**Table S5.** Atoms used to determine the amino-acid side chain CPK volumes with the Spartan '18 software.

| R | Amino acid | Amino-acid side chain CPK volume |
|---|---|---|
| H | Glycine | 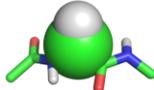 |
| Me | Alanine | 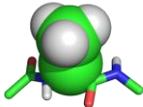 |
| Et | 2-Aminobutiric acid | 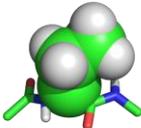 |
| *i*Pr | Valine | 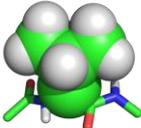 |
| *t*Bu | *tert*-Butylglycine | 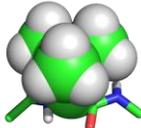 |
| *i*Bu | Leucine | 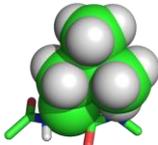 |
| Ph | Phenylglicine | 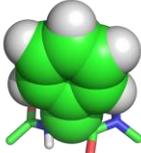 |
| CH$_2$Ph | Phenylalanine | 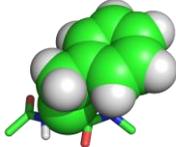 |

The macrocycle cavity was obtained by using the *Cavities & Pockets* function in the PyMOL software (Figure S23a).[S7] The cavity size was calculated using the pywindow software and displayed in PyMOL as a sphere centered in the center of mass of the molecule (Figure S23b).[S8]

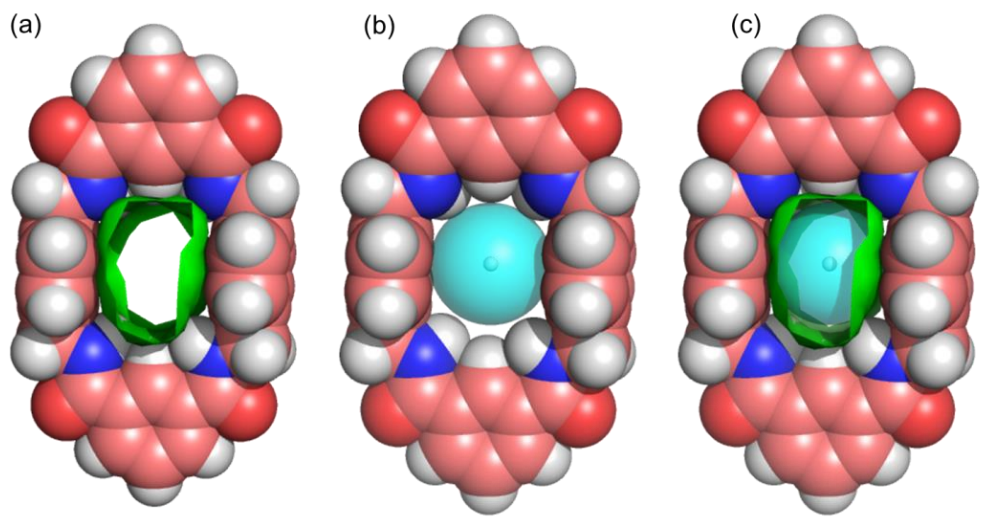

**Figure S23**. Macrocycle structure from Figure S21. (a) Cavity of the macrocycle obtained using the *Cavities & Pockets* function in the PyMOL software. (b) Cavity size represented as a sphere of radius 2.44 Å centered in the center of mass of the molecule calculated using the pywindow software. (c) Overlay of the cavity and sphere calculations.

**Table S6**. Predicted $\Delta G^{\ddagger}$ and $k_{ex}$ in TCE-$d_2$ at 25 ºC using the equation $\Delta G^{\ddagger} = 0.1579 \times V_{CPK} + 5.7227$ from the main manuscript. Experimental data in brackets.

| R | $V_{CPK}$ (Å$^3$) | Predicted $\Delta G^{\ddagger}$ (kcal/mol) | Predicted $k_{ex\ 25\ ºC}$ (s$^{-1}$) |
|---|---|---|---|
| H | 30.54 | 10.5 | $1.1 \times 10^5$ |
| Me | 49.12 | 13.5 (13.4) | 810 (980) |
| Et | 67.38 | 16.4 (16.5) | 6.2 (4.6) |
| *i*Pr | 85.35 | 19.0 (19.1) | 0.054 (0.07) |
| *t*Bu | 102.98 | 22.0 | $4.7 \times 10^{-4}$ |
| *i*Bu | 106.26 | 22.5 | $1.9 \times 10^{-4}$ |
| Ph | 114.92 | 23.9 | $1.9 \times 10^{-5}$ |
| CH$_2$Ph | 133.04 | 26.7 | $1.5 \times 10^{-7}$ |

# NMR spectra

## Rotaxane 1a

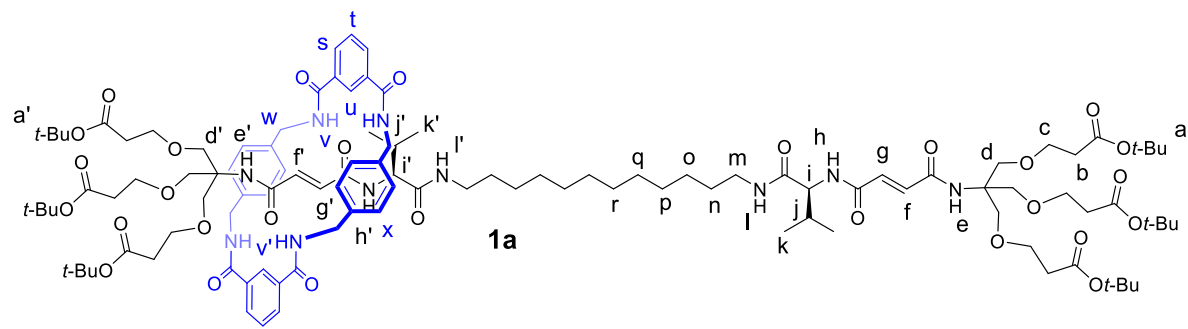

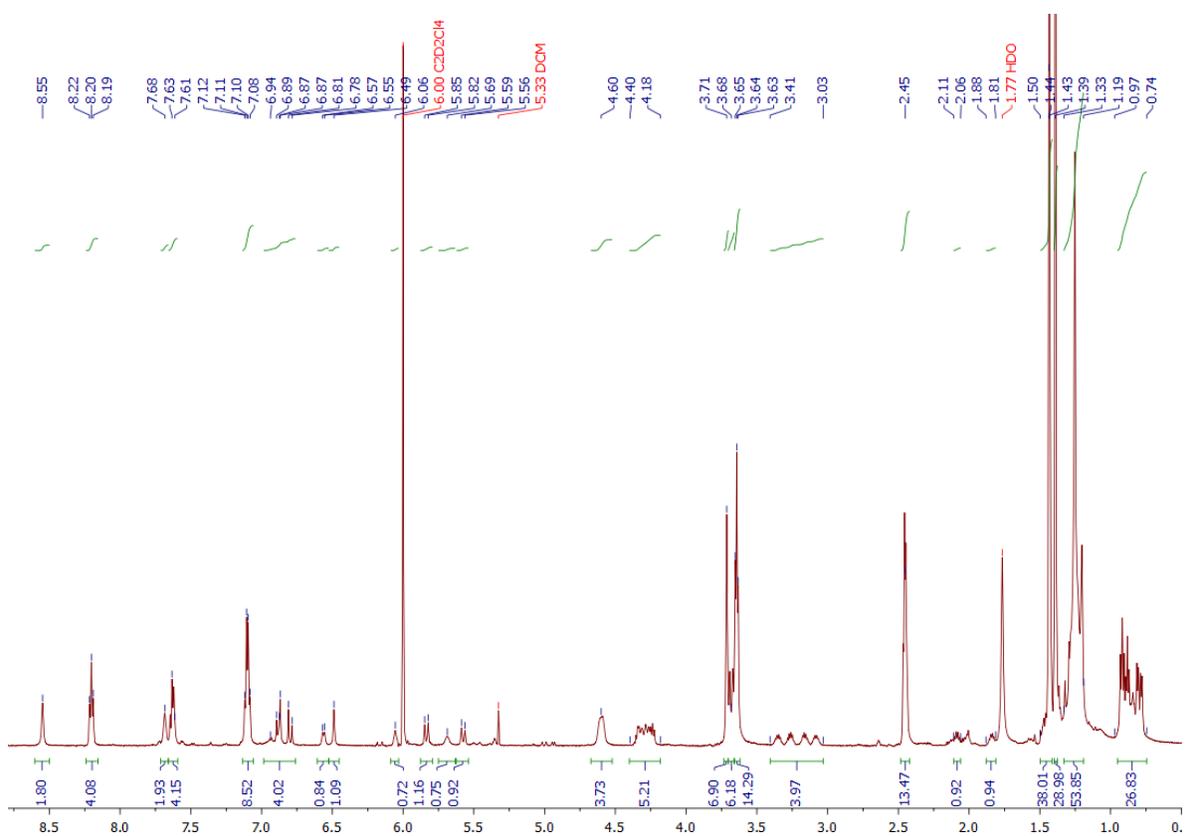

**Figure S24**. ¹H-NMR of valine-containing rotaxane **1a** (600 MHz in TCE-$d_2$).

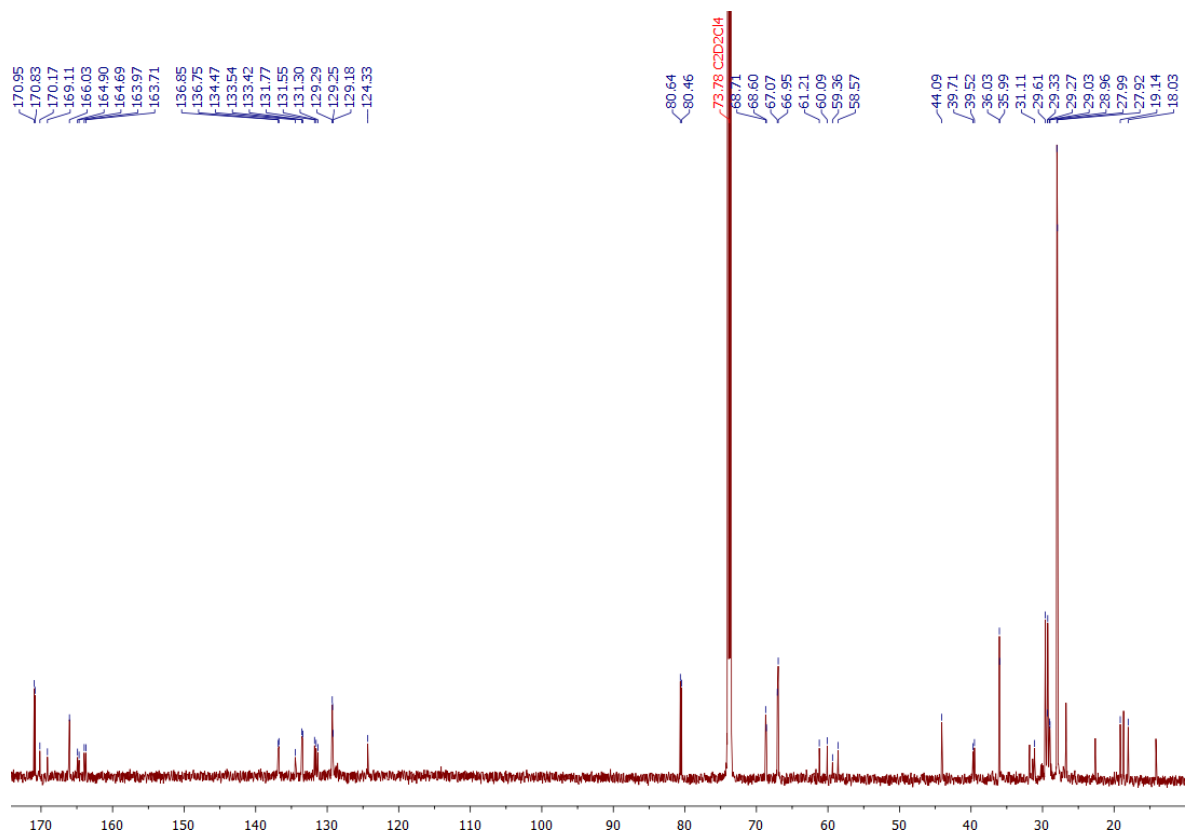

**Figure S25.** $^{13}$C-NMR of valine-containing rotaxane **1a** (151 MHz in TCE-$d_2$).

## Rotaxane **1b**

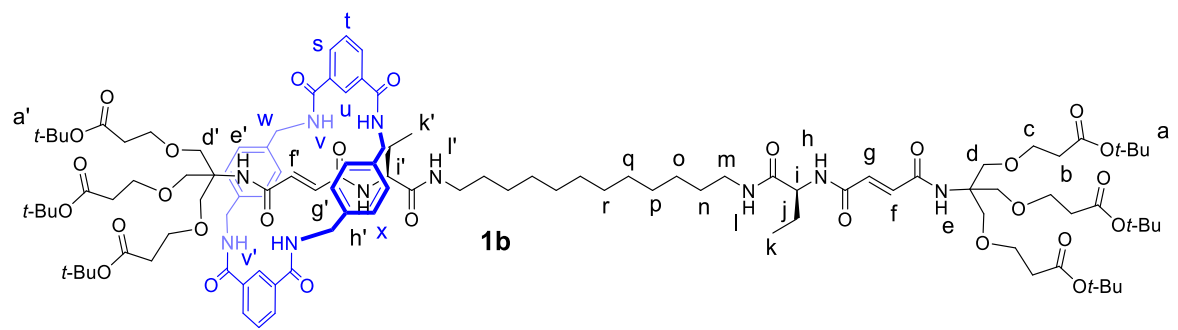
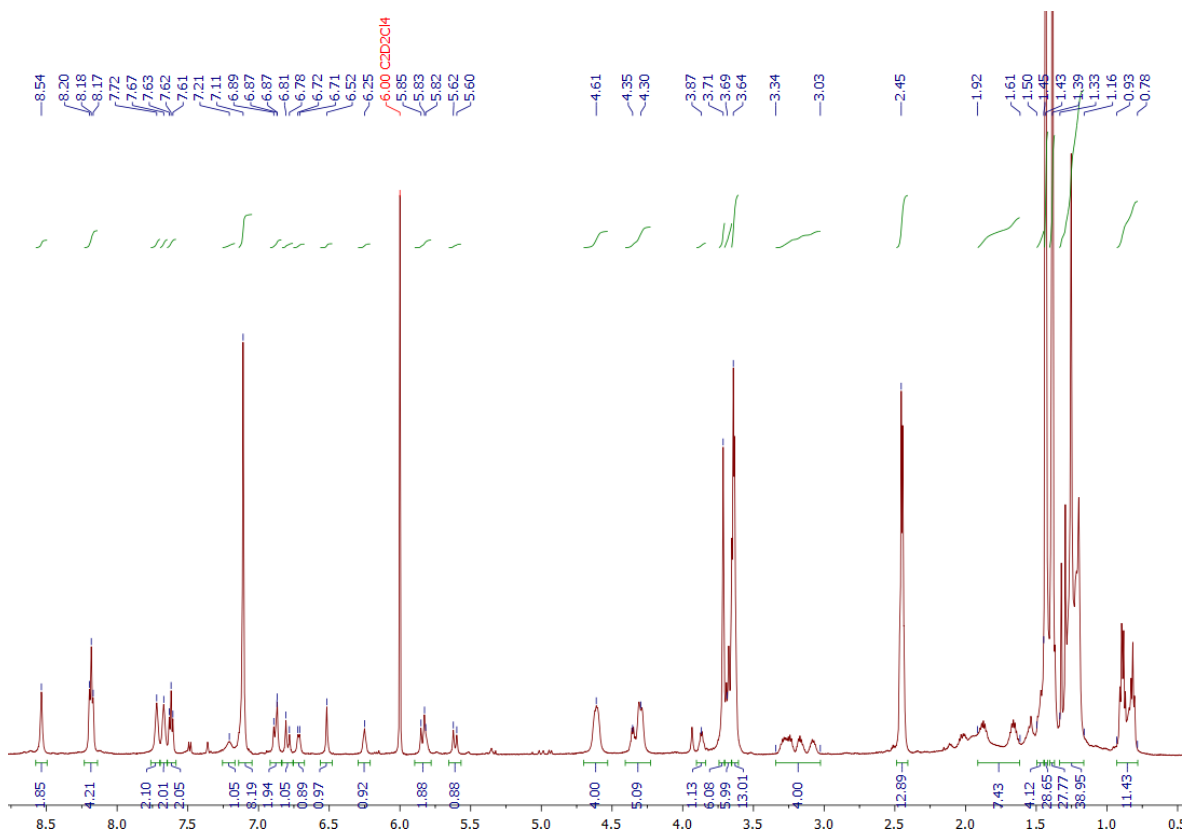

**Figure S26**. ¹H-NMR of 2-aminobutanoic acid-containing rotaxane **1b** (600 MHz in TCE-$d_2$).

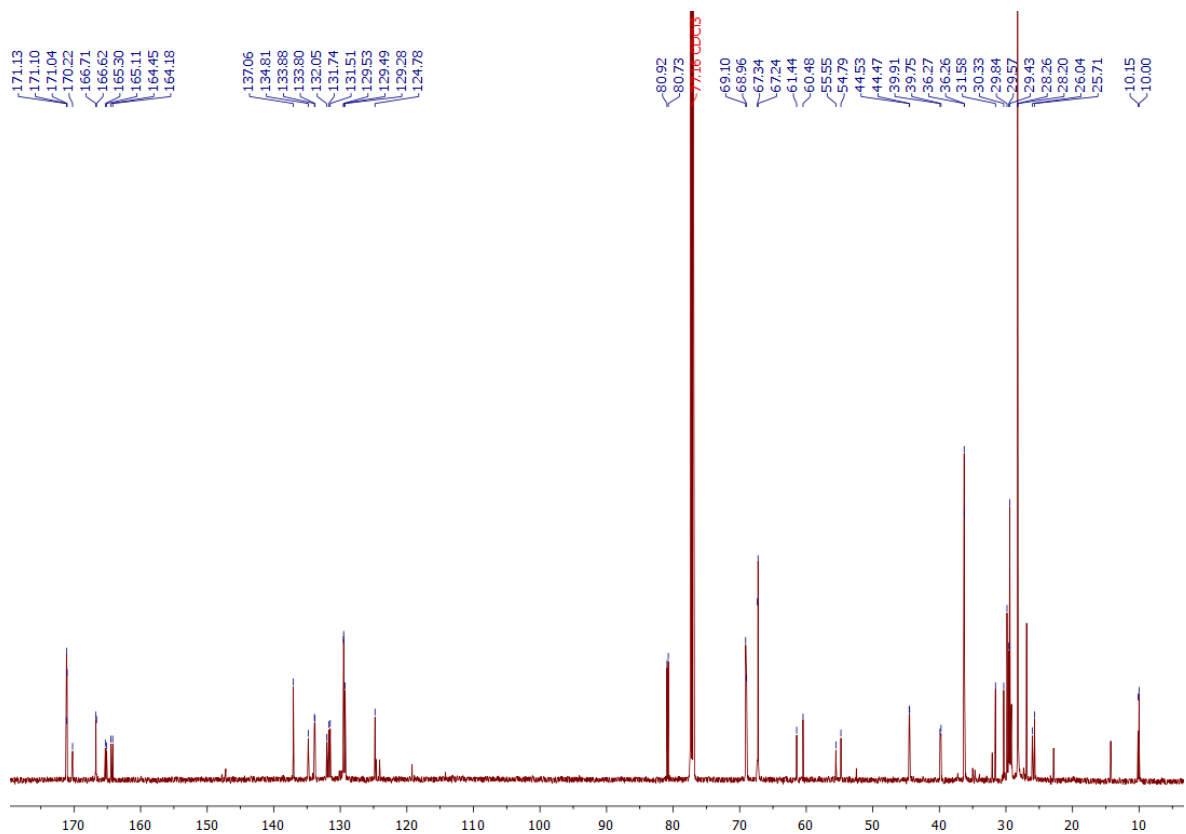

**Figure S27**. $^{13}$C-NMR of 2-aminobutanoic acid-containing rotaxane **1b** (151 MHz in CDCl$_3$).

## Rotaxane **1c**

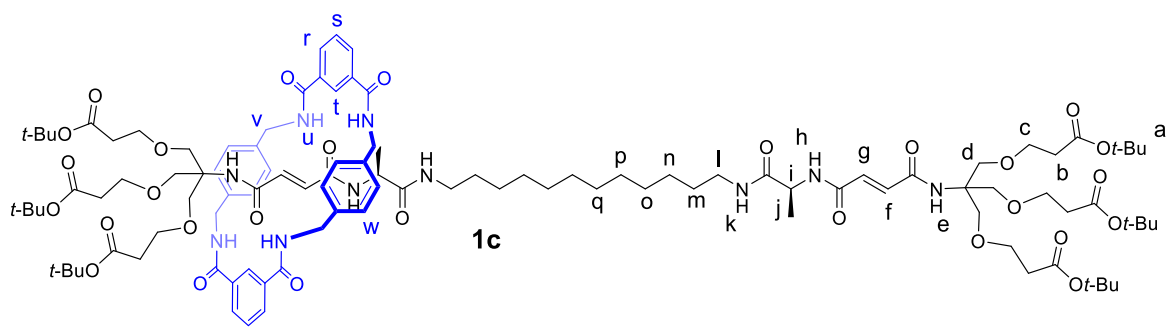

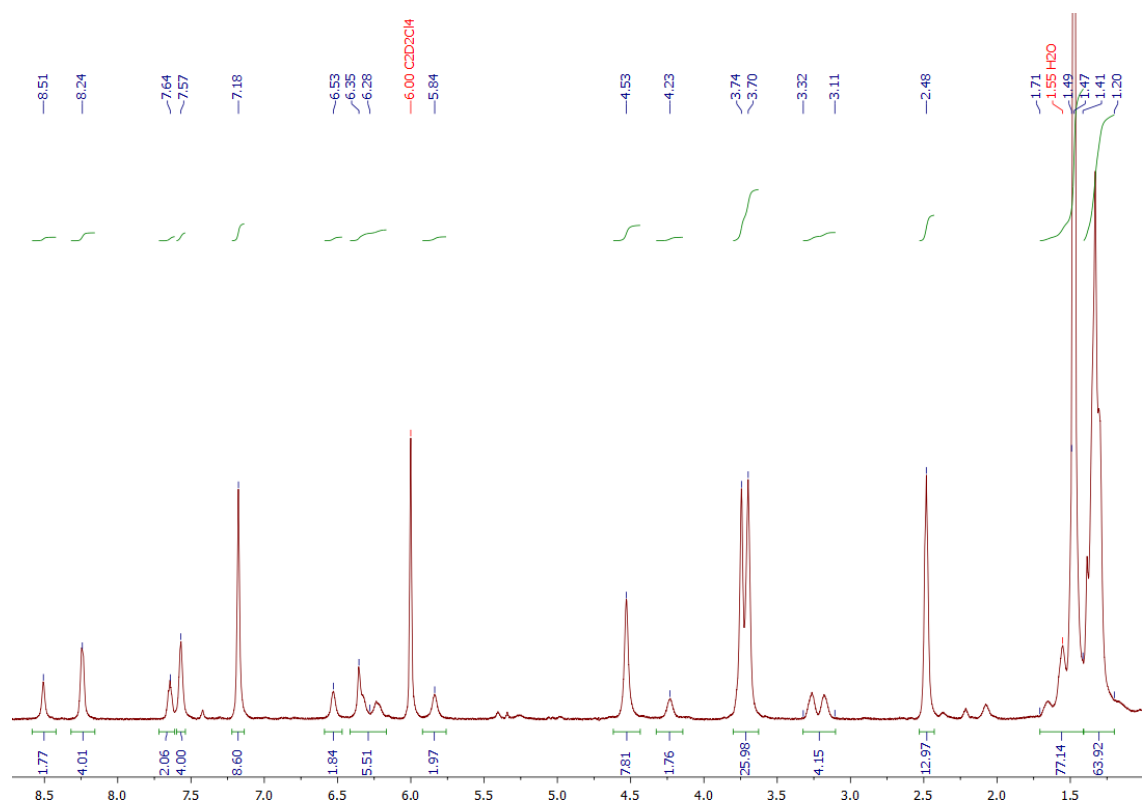

**Figure S28**. ¹H-NMR of alanine-containing rotaxane **1c** (600 MHz in TCE-$d_2$ at 85°C).

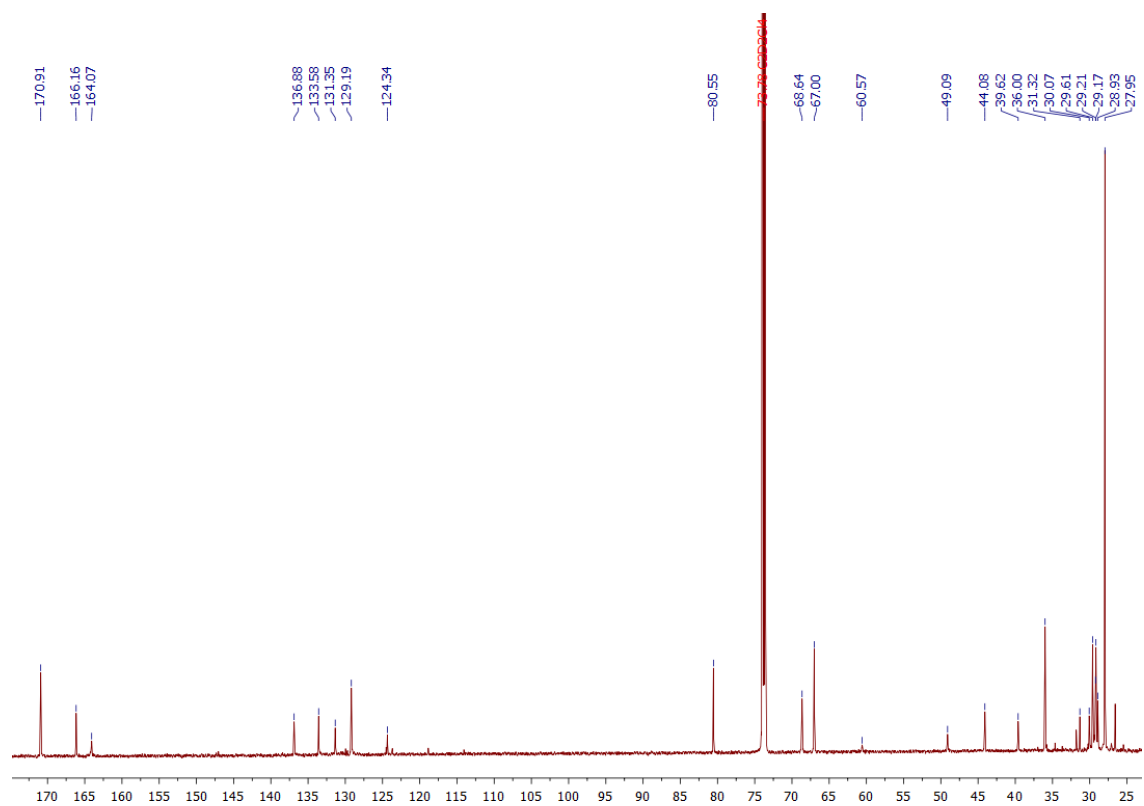

**Figure S29**. $^{13}$C-NMR of alanine-containing rotaxane **1c** (151 MHz in TCE-$d_2$).

## Rotaxane **1a-WS**

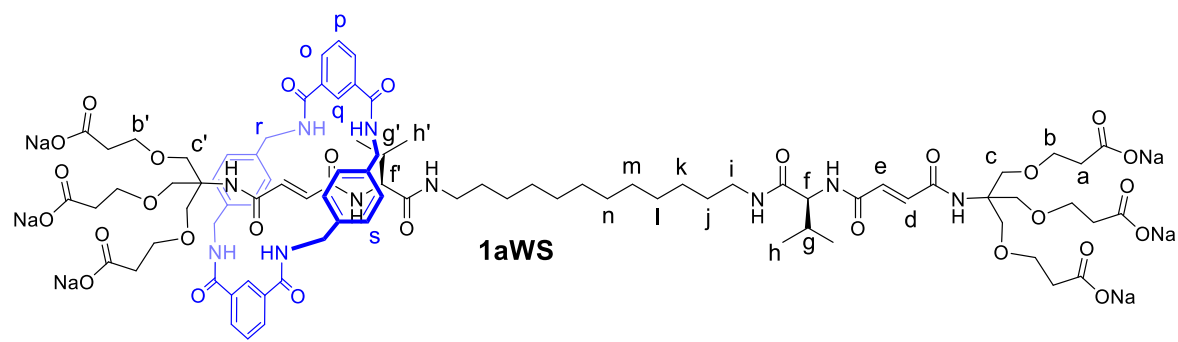

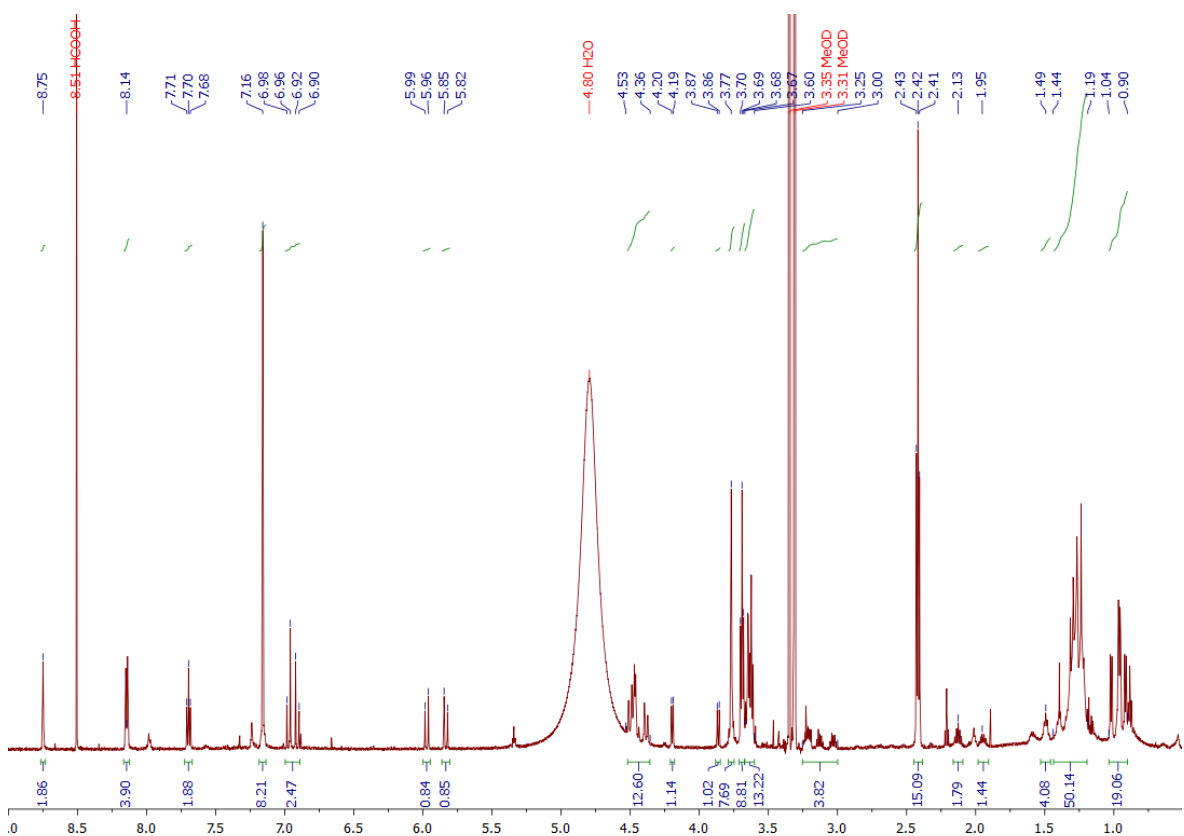

**Figure S30**. ¹H-NMR of valine-containing water soluble rotaxane **1a-WS** (600 MHz in MeOD/D$_2$O 9:1, v/v).

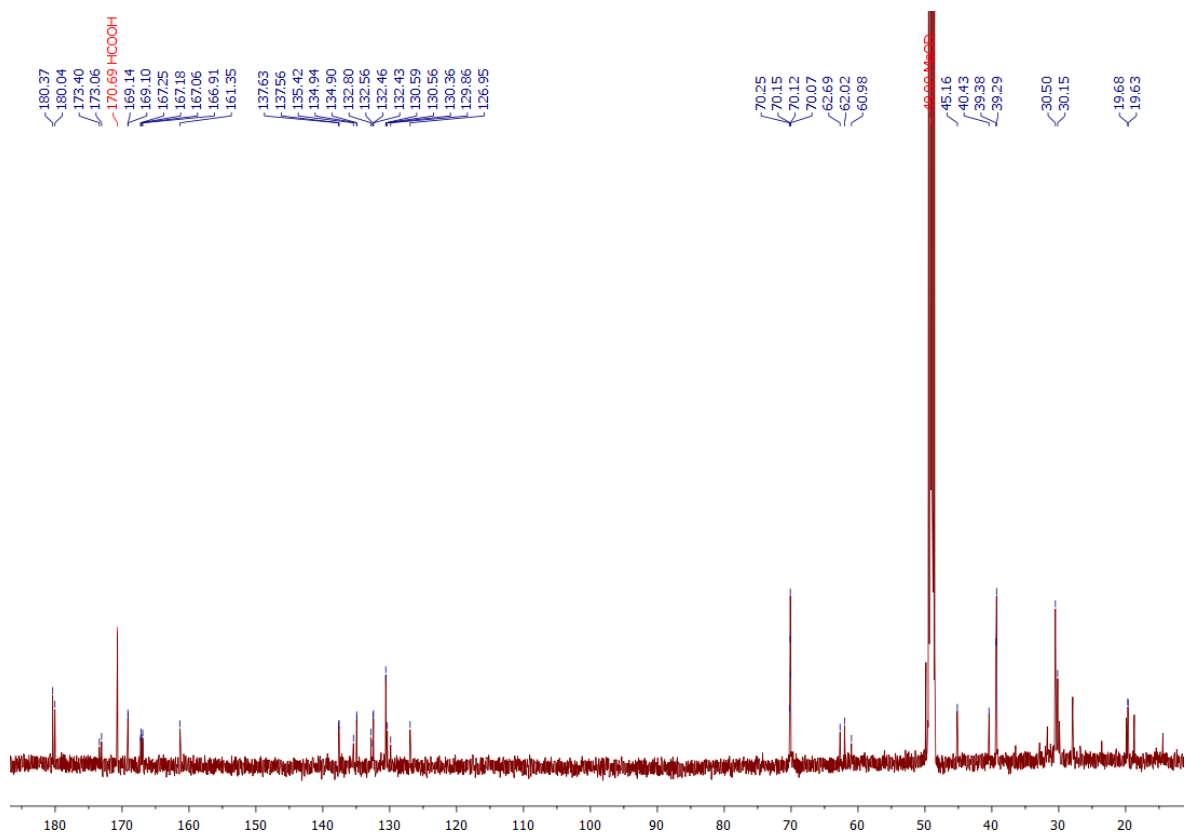

**Figure S31**. $^{13}$C-NMR of valine-containing water soluble rotaxane **1a-WS** (151 MHz in MeOD/D$_2$O 9:1, v/v).

## Compound **2a**

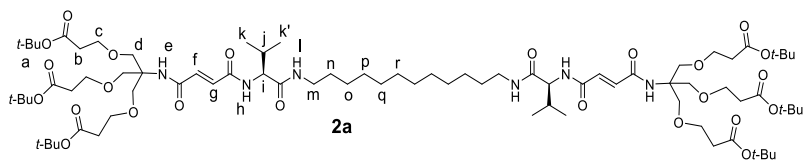

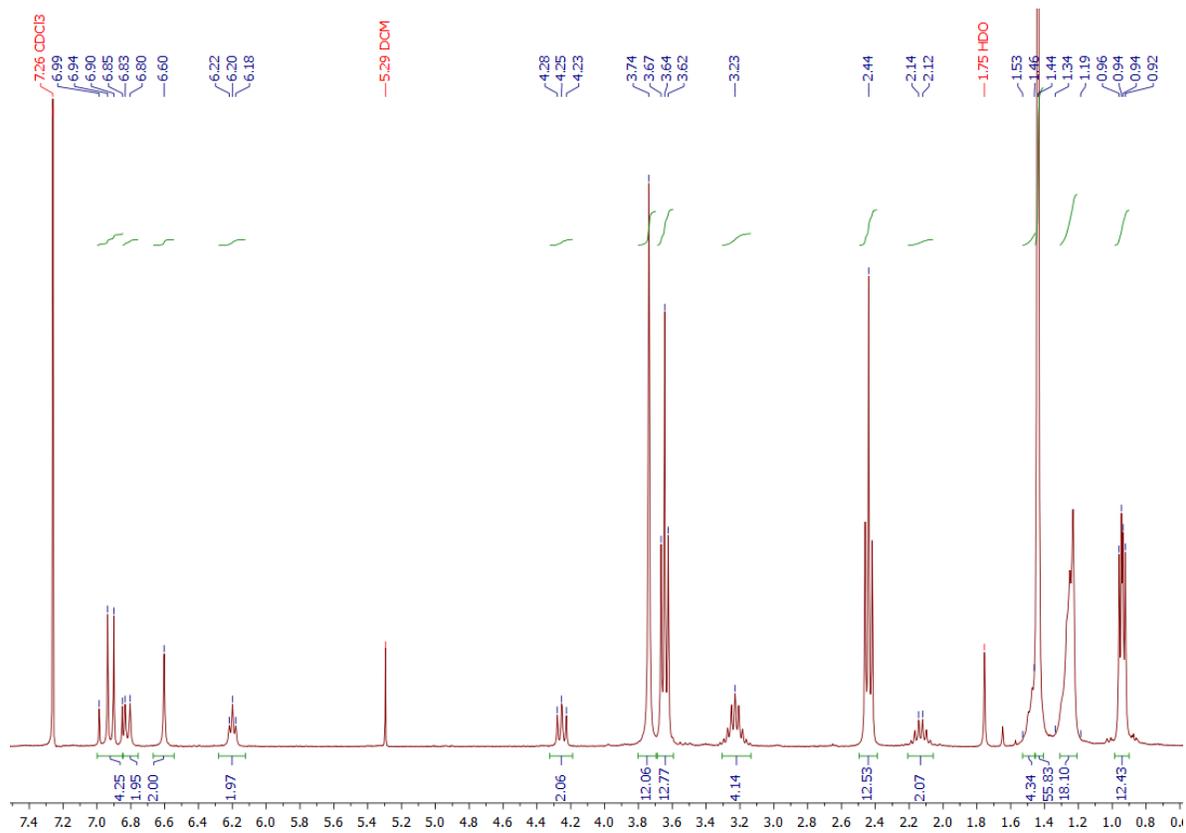

**Figure S32**. [1]H-NMR of valine-containing thread **2a** (300 MHz in CDCl$_3$).

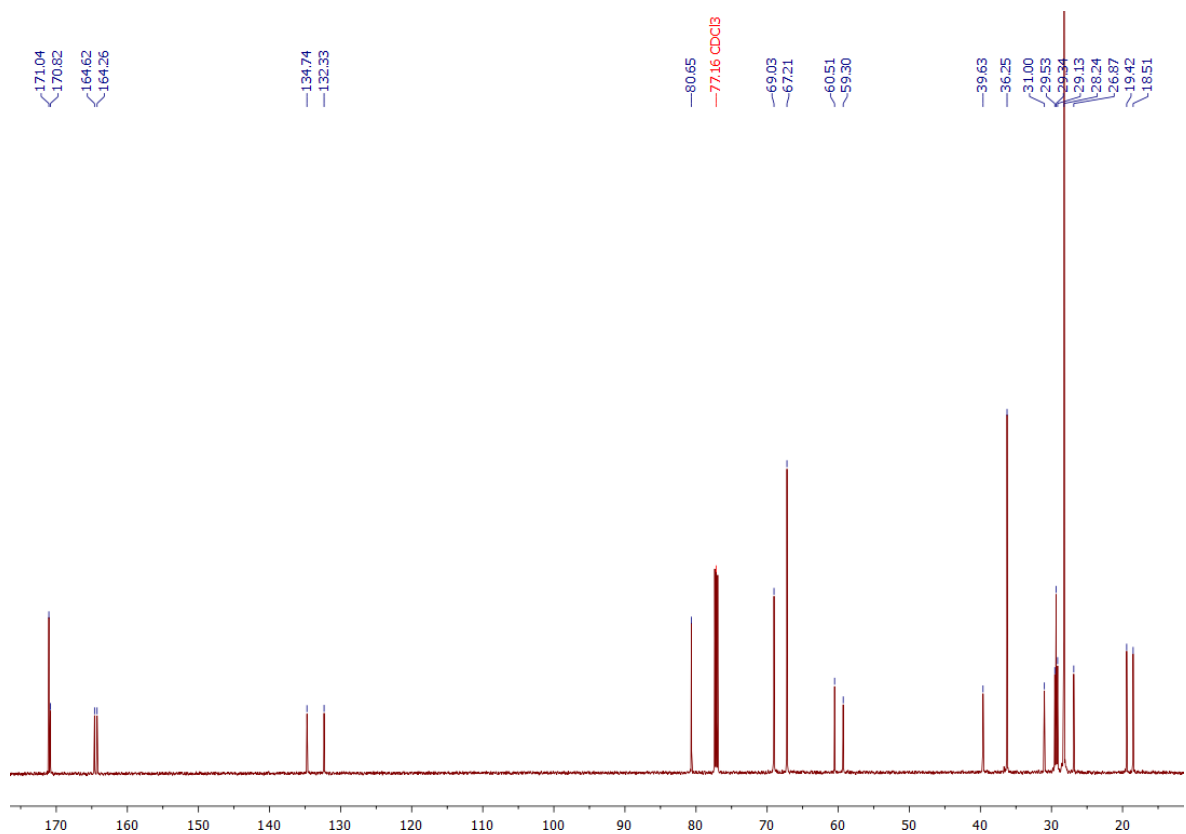

**Figure S33**. $^{13}$C-NMR of valine-containing thread **2a** (151 MHz in CDCl$_3$).

## Compound **2b**

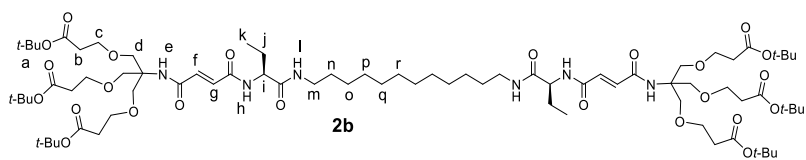

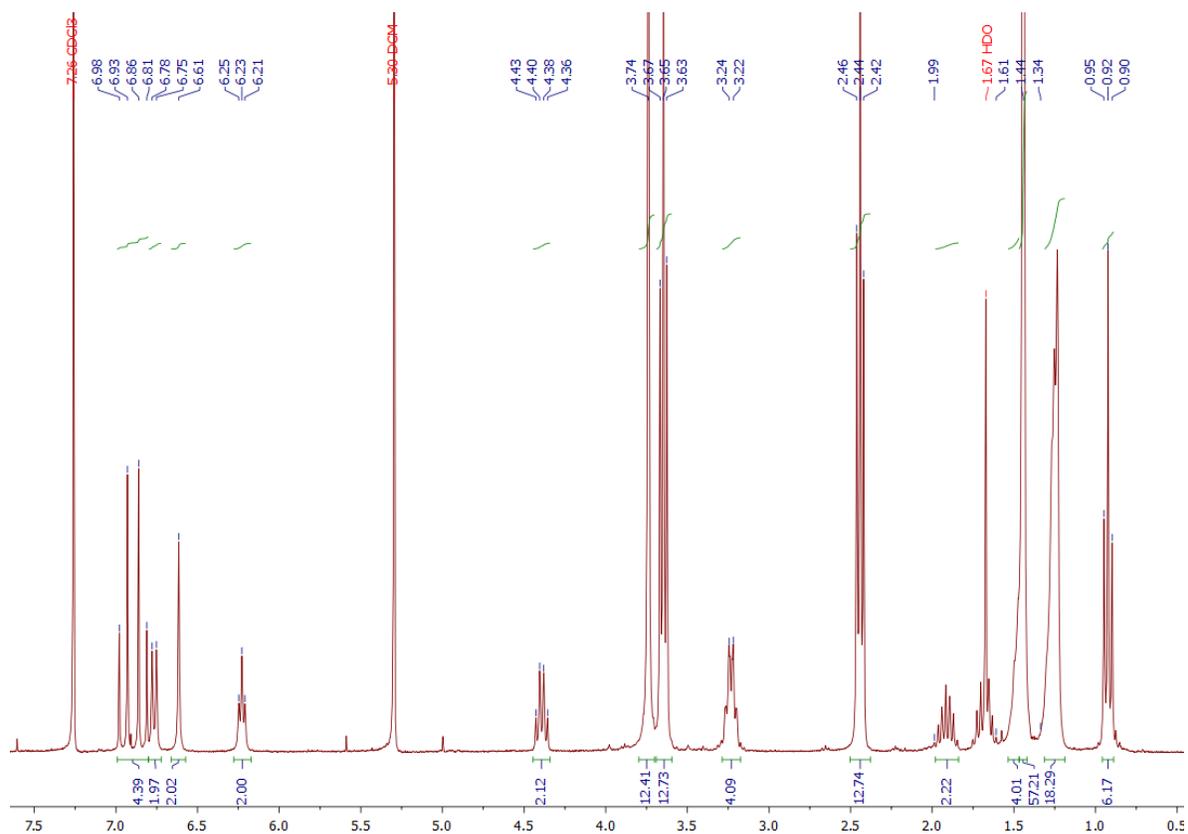

**Figure S34**. ¹H-NMR of 2-aminobutanoic acid-containing thread **2b** (300 MHz in CDCl$_3$).

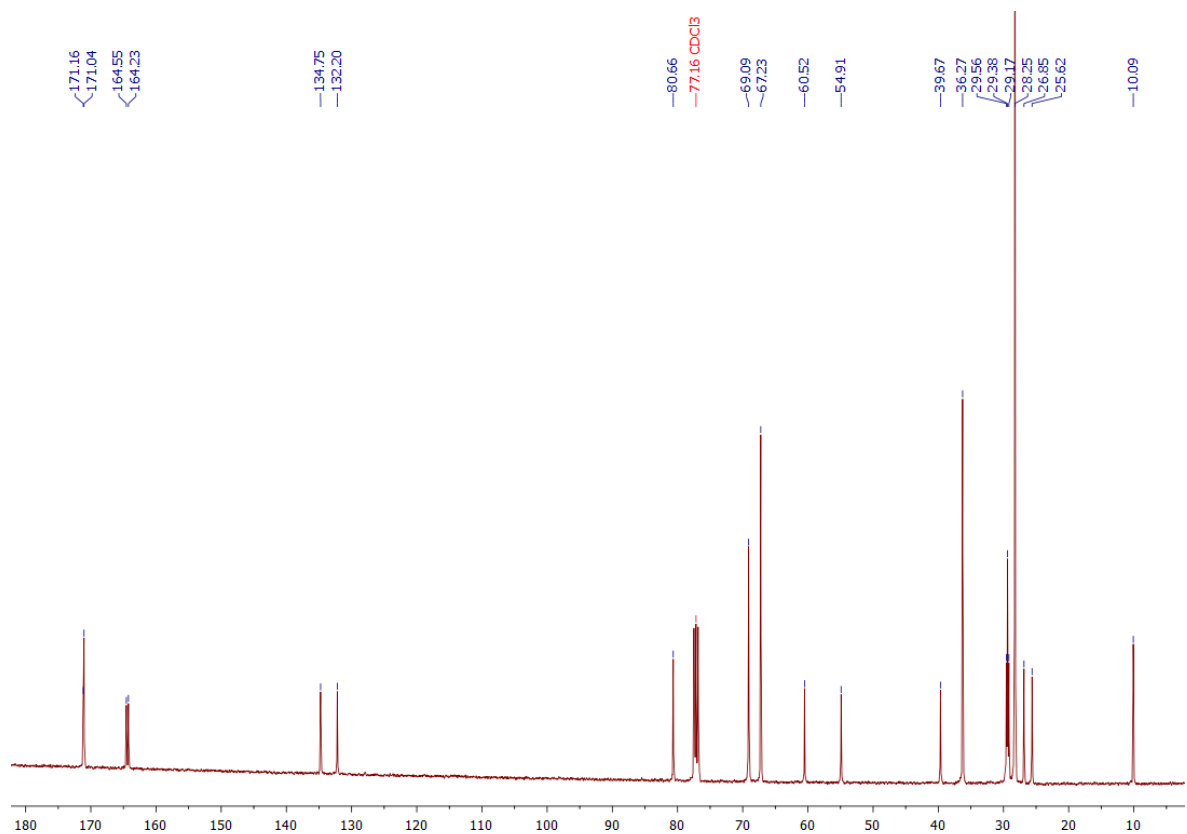

**Figure S35.** $^{13}$C-NMR of 2-aminobutanoic acid-containing thread **2b** (101 MHz in CDCl$_3$).

## Compound **2c**

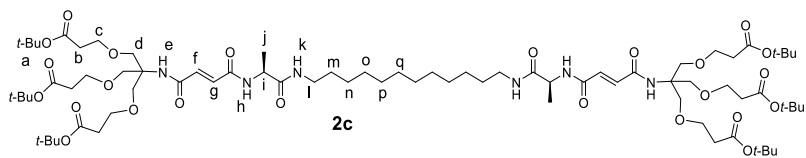

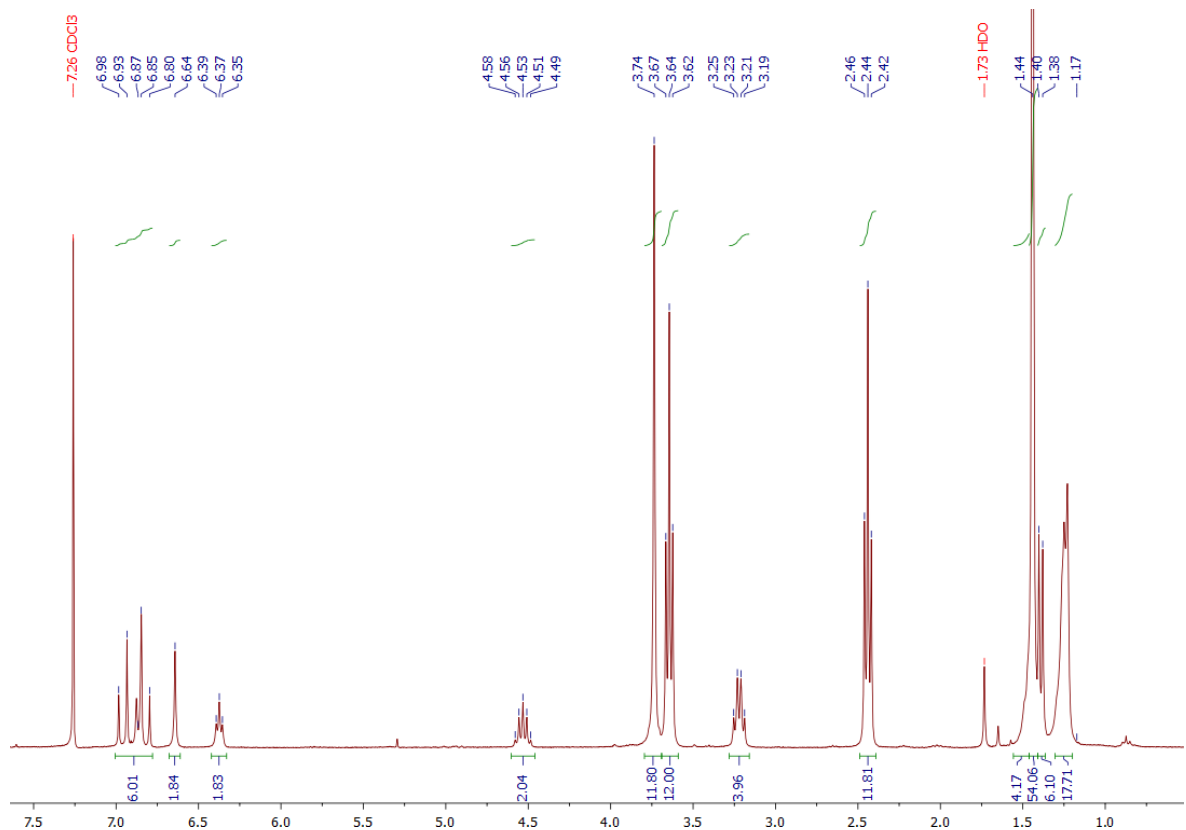

**Figure S36**. ¹H-NMR of alanine-containing thread **2c** (300 MHz in CDCl₃).

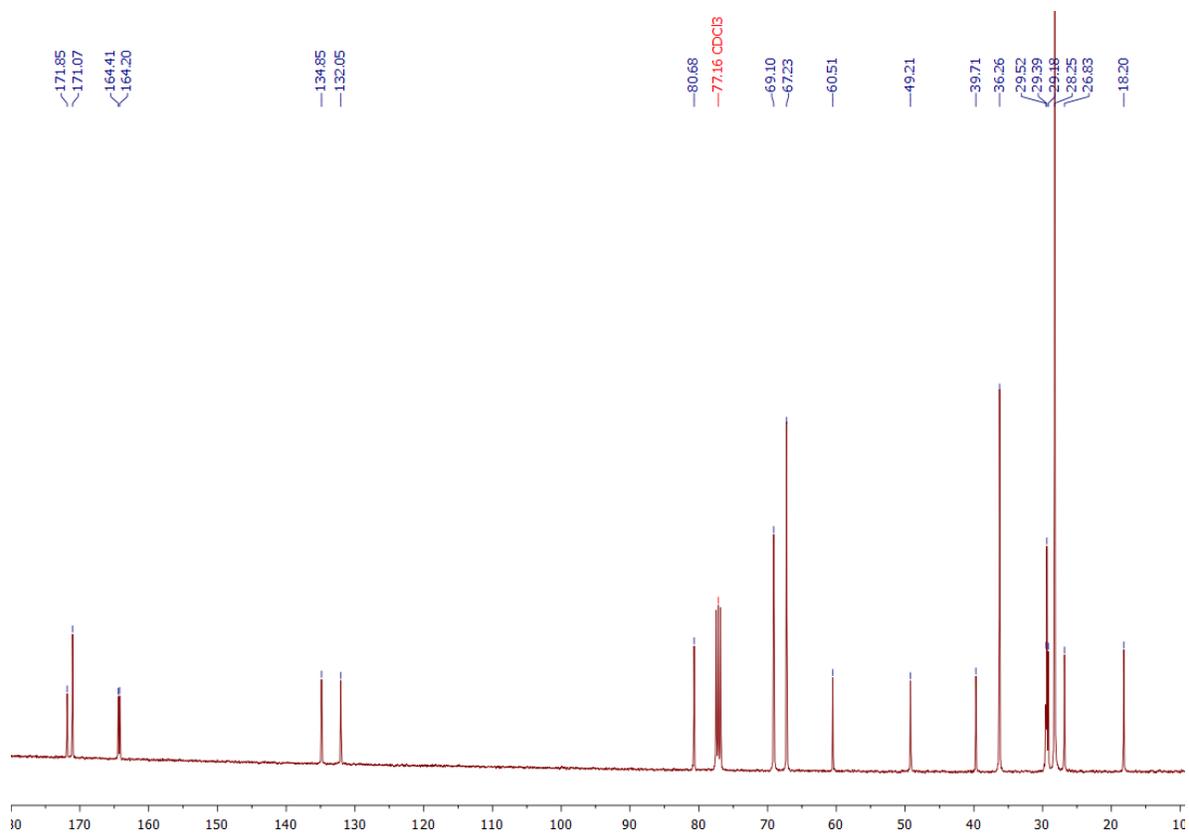

**Figure S37**. $^{13}$C-NMR of alanine-containing thread **2c** (101 MHz in CDCl$_3$).

## Compound 4

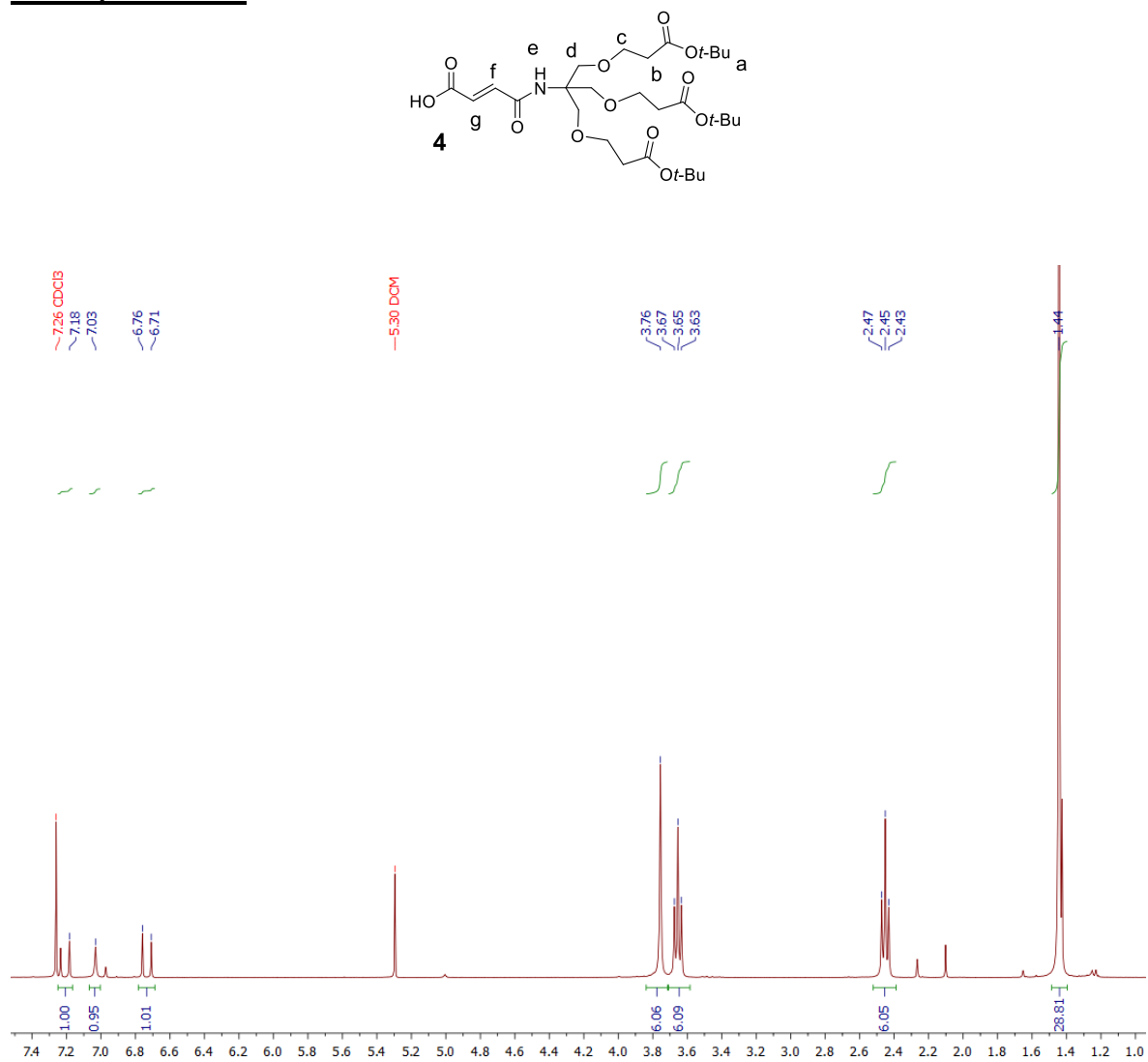

**Figure S38**. $^1$H-NMR of fumaric acid stopper **4** (300 MHz in CDCl$_3$).

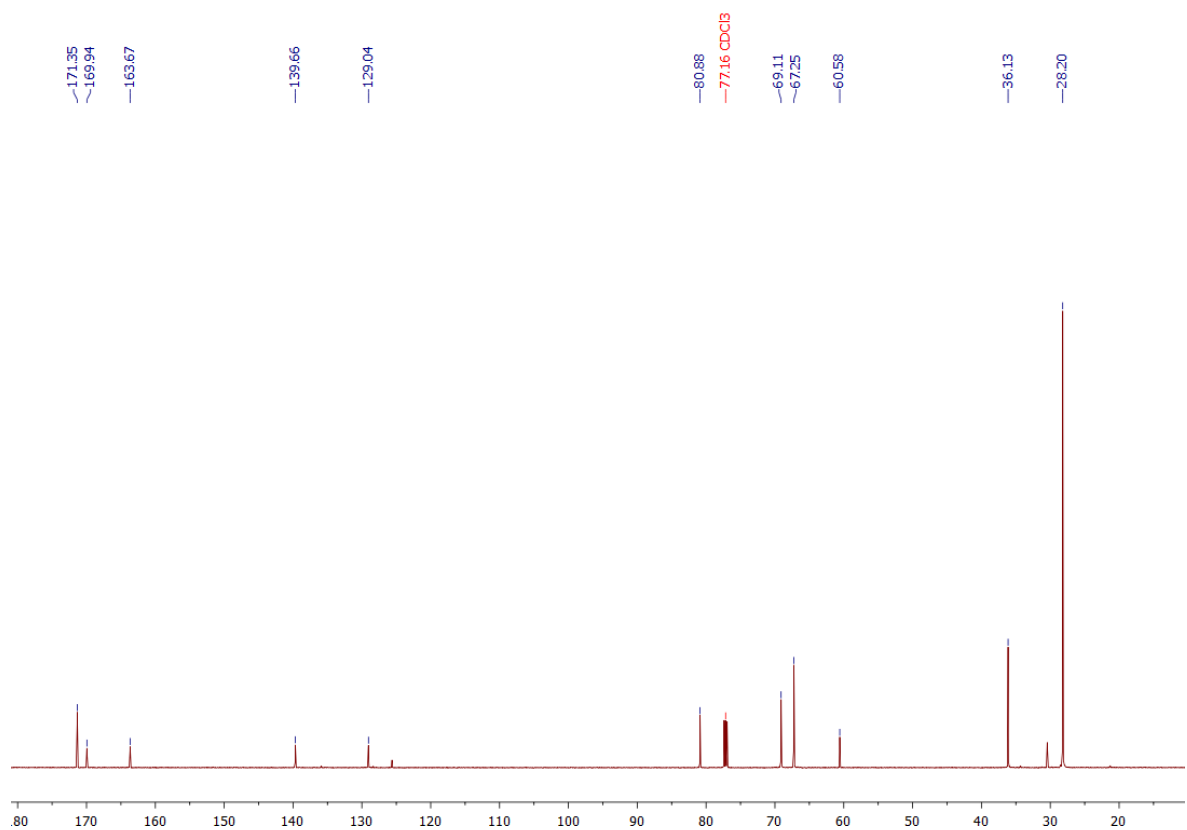

**Figure S39**. $^{13}$C-NMR fumaric acid stopper **4** (151 MHz in CDCl$_3$).

## Compound 5

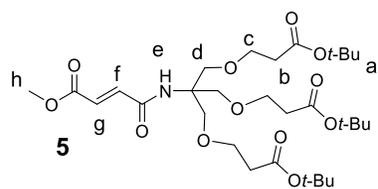

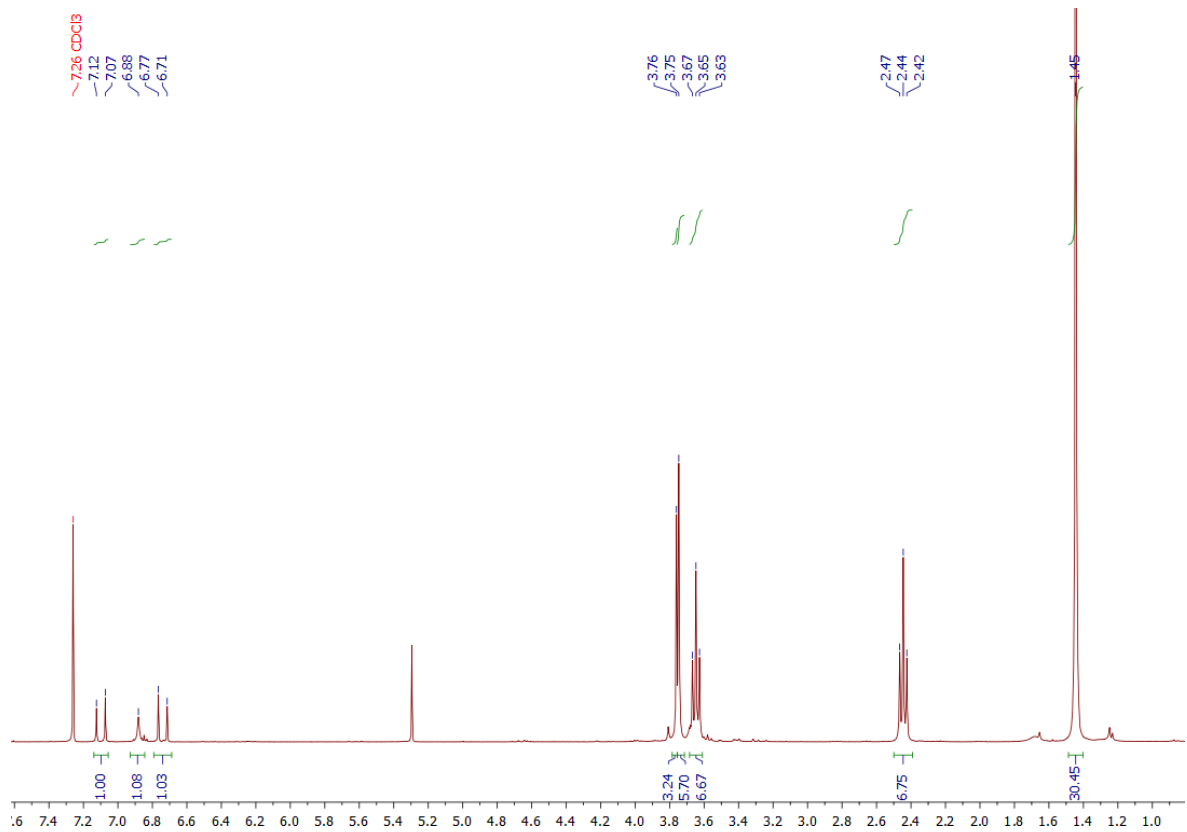

**Figure S40.** $^1$H-NMR of methyl ester **5** (300 MHz in CDCl$_3$).

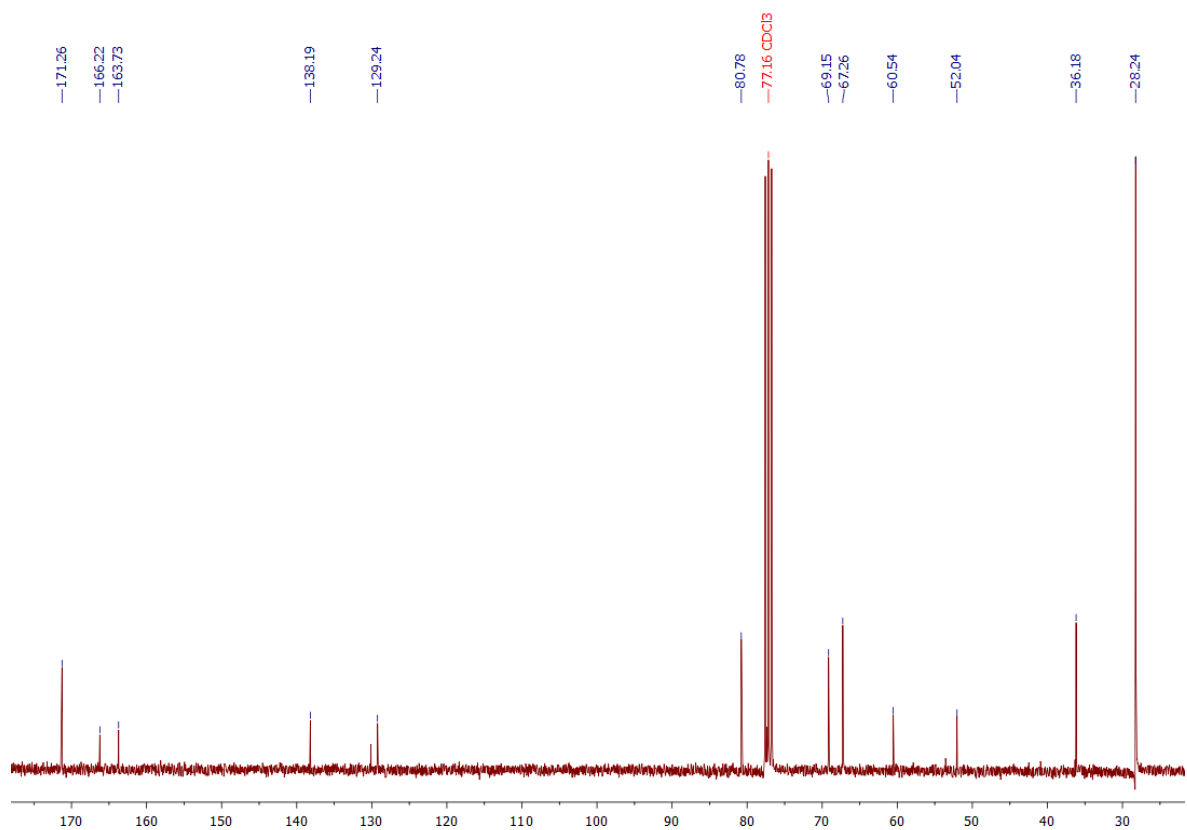

**Figure S41**. $^{13}$C-NMR of methyl ester **5** (76 MHz in CDCl$_3$).